\documentclass[12pt]{article}
\usepackage{jheppub} 

\usepackage{amsmath,amssymb}
\usepackage{url}

\usepackage{graphicx}
\usepackage{caption}
\usepackage{subcaption}
\usepackage[utf8]{inputenc}
\usepackage[T1]{fontenc}

\usepackage{color}
\input{colordvi.tex}

\usepackage{hyperref}

\def\be{\begin{eqnarray}}
\def\ne{\nonumber\end{eqnarray}}
\def\ee{\end{eqnarray}}

\def\c{\cdot}

\def\sitarel#1#2{\mathrel{\mathop{\kern0pt #1}\limits_{#2}}}
\def\={\buildrel\bigtriangledown\over=}
\def\!={\buildrel!\over=}
\def\?={\buildrel?\over=}

\def\a{\alpha}

\newcommand{\colvec}[1]{\left(\begin{array}{c}#1\end{array}\right)}

\newcommand{\colmatt}[1]{\left(\begin{array}{cc}#1\end{array}\right)}

\begin{document}

\title{Phase Transitions at Unusual Values of $\theta$}

\author[a]{Csaba Cs\'aki,}
\author[b]{Teruhiko Kawano,}
\author[c,d,e]{Hitoshi Murayama}
\author[f]{and Ofri Telem}

\affiliation[a]{Department of Physics, LEPP, Cornell University, Ithaca, NY 14853, USA}
\affiliation[b]{Department of Physics, Kyoto Prefectural University of Medicine, Kyoto 606-0823, Japan}
\affiliation[c]{Department of Physics, University of California, Berkeley, CA 94720, USA}
\affiliation[d]{Kavli Institute for the Physics and Mathematics of the
  Universe (WPI), University of Tokyo,
  Kashiwa 277-8583, Japan}
\affiliation[e]{Ernest Orlando Lawrence Berkeley National Laboratory, Berkeley, CA 94720, USA}
\affiliation[f]{Racah Institute of Physics, Hebrew University of Jerusalem, Jerusalem 91904, Israel}

\emailAdd{csaki@cornell.edu}
\emailAdd{kawano@koto.kpu-m.ac.jp}
\emailAdd{hitoshi@berkeley.edu}
\emailAdd{t10ofrit@gmail.com}

\vskip .5in 
\abstract{
We calculate the $\theta$ dependence in a cousin of QCD, where the vacuum structure can be analyzed exactly. The theory is $\mathcal{N}=2$ $SU(2)$ gauge theory with $N_F=0,1,2,3$ flavors of fundamentals, explicitly broken to $\mathcal{N}=1$ via an adjoint superpotential, and coupled to anomaly mediated supersymmetry breaking (AMSB). The hierarchy $m_{AMSB}\ll \mu_{\mathcal{N}=1}\ll \Lambda$ ensures the validity of our IR analysis. As expected from ordinary QCD, the vacuum energy is a function of $\theta$ which undergoes  1st order phase transitions between different vacua where the various
dyons condense. For $N_F=0$ we find the expected phase transition at $\theta=\pi$, while for $N_F=1,2,3$ we find phase transitions at fractional values of $\pi$.
}
\maketitle
\flushbottom

\section{Introduction}

The $\theta$ dependence of the vacuum  is one of the most important and difficult aspects of the strong/confining dynamics in gauge theories. Even within real world QCD this question is not fully clarified, including the dynamical origin of the various terms appearing in the potential. { Initially it was assumed that the $\theta$-dependence is due to instanton effects, which seems like a reasonable guess since instantons are intimately related to the existence of the $\theta$ vacua. However Witten~\cite{Witten:1978bc} and di Vecchia and Veneziano~\cite{DiVecchia:1980yfw} realized that this is unlikely (see also \cite{Ohta:1981ai} for a complementary study). The essence of the argument is as follows: when fermionic matter is introduced, an anomalous chiral symmetry appears which rotates the $\theta$ term. The corresponding Goldstone boson (called $\eta'$) picks up a mass from the same dynamics that is responsible for the $\theta$-dependence of QCD. However since in the large-$N$ limit the $\eta'$ mass has to vanish, the $\theta$-dependent part of the potential has to be more complicated than a simple instanton generated $\Lambda^4 \cos \theta$ term. In particular, it should have several branches, giving rise to first order phase transitions as one moves from on branch to the other. A separate argument for a phase transition in pure Yang-Mills theories was put forward long ago~\cite{Dashen:1970et,tHooft:1981bkw} and first realized in \cite{Cardy:1981qy}. The conjecture was that the} charge-parity (CP) symmetry is spontaneously broken at $\theta=\pi$ in pure Yang-Mills theories, leading to a doubly-degenerate vacuum and indicating a first-order phase transition. Recently, this assertion has been greatly sharpened for $SU(N)$ Yang-Mills (YM) theory in \cite{Gaiotto:2017yup}. The main tool employed in the latter analysis was the identification of a mixed anomaly between the $Z_N$ 1-form center symmetry of $SU(N)$ and time reversal at $\theta=\pi$. 't Hooft anomaly matching then implies one of the following scenarios at $\theta=\pi$: (a) the theory is gapless; (b) the theory is non-trivially gapped and the IR topological field theory reproduces the anomaly; (c) the $Z_N$ 1-form center symmetry is spontaneously broken and so the theory does not confine; or (d) time reversal is spontaneously broken, leading to a doubly degenerate vacuum and a first order phase transition at $\theta=\pi$. While none of the above scenarios is ruled out, option (d) was seen as the most probable - and the only one consistent with the analysis of softly broken SUSY gauge theory and with large-$N$ studies.

The non-perturbative nature of the IR dynamics in asymptotically free theories makes the study of the $\theta$-dependence particularly hard, and one has to either specialize to supersymmetric 
\cite{Veneziano1982,Taylor1983,Affleck:1983vc,Affleck:1983mk,Affleck1985,Seiberg1994a,Seiberg1994,Seiberg:1994bz,Seiberg1995,Douglas1995,Hanany1995,Intriligator:1995id,Leigh1995,Pouliot:1995me,Kutasov1995,Finnell:1995dr,Kutasov1995a,Intriligator1995,Argyres1995,Argyres1995a,Argyres1995b,Matone1995,Klemm1995,Murayama:1995ng,Klemm1996,Argyres1996,Argyres1996a,Donagi1996,Intriligator1996,Kutasov1996,Pouliot1996,Pouliot1996a,Intriligator1996a,Aharony1997,Cachazo2002,Nekrasov2003,Auzzi2003,Intriligator2006,Pestun2012} or near-supersymmetric gauge theories \cite{Aharony:1995zh,Evans:1995ia, DHoker:1996xdz,AlvarezGaume1996,Konishi1997,Evans1997,AlvarezGaume1997,AlvarezGaume1998,AlvarezGaume1998a,Cheng1998,Martin:1998yr,ArkaniHamed1998,Luty1999,Abel2011,Cordova2018,Csaki:2021xhi,Csaki:2021aqv,Csaki:2021jax,Murayama:2021xfj,Csaki:2021xuc,Kondo:2021osz,Luzio:2022ccn,Csaki:2022cyg,Dine:2022req,Dine:2022nmt} or make use of consistency conditions like anomaly matching \cite{Raby:1979my,Hooft1980,Dimopoulos:1980hn,Csaki:1997aw,Gaiotto:2017yup,Tanizaki2017,Gaiotto2018,Cordova2018,Shimizu2018,Bi2019,Tanizaki:2018wtg,Tanizaki2018,Hsin2019,GarciaEtxebarria2019,Cordova2020,Freed2021,Cordova2020a,Bolognesi:2020mpe,Bolognesi:2021yni,Tong2020,Razamat2021,Karasik2022,Smith2022}, or of large-$N$ arguments.  
In this paper, we perform a complementary study to the one in \cite{Gaiotto:2017yup}, by analyzing a particular set of near-supersymmetric theories, where the vacuum energy can be calculated exactly as a function of $\theta$. Our intention is the study of this particular toy model rather than the extrapolation to any particular non-supersymmetric gauge theory (e.g. QCD). For this reason, we are free to choose our original supersymmetric gauge theory as well as our method of SUSY breaking. The validity of our analysis is guaranteed as long as we correctly map the SUSY breaking from the UV to the IR, and as long as we maintain a hierarchy $m_{breaking}\ll \Lambda$, the strong scale of the theory. 

Our theory of interest is $\mathcal{N}=2$ gauge theory with gauge group $SU(2)$ and $N_F=0,1,2,3$ flavors of fundamentals. {This is particularly suitable for studying the phases of pure QCD-like theories. The reason is that adding matter fields will usually introduce an $\eta'$ into the spectrum, which will then act as a heavy axion and wash out the branches and phase structure of the pure QCD-like theory (see~\cite{Csaki:2023yas} for a new analysis of the $\eta'$ potentials in such models). However $\mathcal{N}=2$ theories with matter have the tree-level $\tilde{Q}\Phi Q$ coupling in their superpotential, which explicitly breaks the usual axial symmetry  and eliminates the relation between $\eta'$ and $\theta$.\footnote{The $\tilde{Q}\Phi Q$ term does preserve a chiral $U(1)$ symmetry under which $Q,\tilde{Q}$ have charge $1$ and $\Phi$ has charge $-2$. However this symmetry will be broken when ${\cal N}=2$ is broken to ${\cal N}=1$ via the adjoint mass term, and this breaking sets the resulting $\eta'$ to zero, removing it from the dynamics and uncovering the QCD-like branch structure for the $\theta$ dependence.} Thus studying $\mathcal{N}=2$ with different $N_F$ and supersymmetry breaking is an ideal tool for understanding the dynamics leading to the phase structure of QCD-like theories. These $\mathcal{N}=2$ theories are the ones explored} in the groundbreaking work of Seiberg and Witten~\cite{Seiberg1994,Seiberg1994a}. In the IR, the theory has an $\mathcal{N}=2$ supersymmetric Coulomb branch parametrized by the vacuum expectation value (VEV) of the adjoint scalar in the $\mathcal{N}=2$ gauge multiplet. On the Coulomb branch, the gauge symmetry is higgsed to its $U(1)$ subgroup. At particular points on the Coulomb branch, the gauge coupling becomes singular, indicating the appearance of new massless degrees of freedom - the monopoles and dyons of the $SU(2)\rightarrow U(1)$ gauge theory. 
As pointed out in the original papers \cite{Seiberg1994,Seiberg1994a}, adding an explicit breaking term of magnitude $\mu\ll \Lambda$ to $\mathcal{N}=1$ lifts the Coulomb branch of the theory while keeping only the monopole/dyon singularities as the true vacua of the IR $\mathcal{N}=1$ theory. Furthermore, the explicit breaking term leads to the condensation of the monopoles/dyons in their respective vacua. At this point, the theory is still $\mathcal{N}=1$ supersymmetric, and so, these vacua are degenerate and have zero vacuum energy. In \cite{Konishi1997,Evans1997}, an additional soft SUSY breaking gaugino mass $m\ll \mu\ll \Lambda$ was introduced in the $N_F=0$ scenario. This had the triple effect of (a) breaking the degeneracy between the different dyon vacua; (b) moving the vacua away from the monopole points by a tiny $\mathcal{O}(m/\Lambda)$ and (c) making the $\theta$ angle physical since all massless fermions are lifted. In particular, the authors of \cite{Konishi1997,Evans1997} noticed a first-order phase transition at $\theta=\pi$ between the monopole and the dyon vacua of the theory, consistent with Dashen's argument and the much later analysis of \cite{Gaiotto:2017yup}. Since then, the systematic study of softly broken $\mathcal{N}=2$ theories has been greatly refined in, e.g. \cite{AlvarezGaume1996,AlvarezGaume1997,AlvarezGaume1998,AlvarezGaume1998a,Luty1999,Cordova2018}, albeit without any particular focus on the $\theta$ dependence of the vacuum energy. See also complementary studies of the theta angle dependence of the vacuum energy in the context of MQCD \cite{Barbon1998,Oz1998}, and large-$N$ \cite{Witten1998,DelDebbio2002,Dine2017}.

The study reported here generalizes the results of \cite{Konishi1997,Evans1997} to $N_F=1,2,3$. Rather than adding a soft mass ``by hand", we make use of the exact techniques developed in \cite{Luty1999} to map the soft SUSY breaking from the UV to the IR. In particular, we make use of anomaly-mediated SUSY breaking \cite{Randall:1998uk,Giudice:1998xp} (see also \cite{ArkaniHamed1998,ArkaniHamed1999} for earlier work containing some important aspects of AMSB), which provides a particularly transparent and streamlined calculation of the IR effects of SUSY breaking. We call a SUSY theory coupled to parametrically small AMSB a \textit{near SUSY} gauge theory. Previously, some of the present authors applied AMSB for the study of $SU(N)$ gauge theory \cite{Murayama:2021xfj,Kondo:2021osz,Csaki2022b}, chiral gauge theories \cite{Csaki:2021xhi,Csaki:2021aqv,Kondo:2022lvu,Leedom:2025mcg}, and $SO(N)$ gauge theory \cite{Csaki:2021jax,Csaki:2021xuc}. See also \cite{Bai:2021tgl,Luzio:2022ccn,Dine:2022req,Dine:2022nmt,deLima:2023ebw}. As we explicitly checked in section~\ref{sec:comp}, breaking SUSY with AMSB is equivalent to choosing a subset of soft parameters in the general formalism of \cite{Luty1999}.

The main result of our paper is the calculation of the vacuum energy of near-SUSY $SU(2)$ gauge theory with $N_F=0,1,2,3$, as a function of $\theta$. For $N_F=0$ we reproduce the expected phase transition at $\theta=\pi$, while for $N_F=1$, we find a surprising additional phase transition at $\theta=0$. For $N_F=2$ we find phase transitions at $\theta=\frac{\pi}{2},\,\frac{3\pi}{2}$, while for $N_F=3$ we find them at $\theta=\frac{\pi}{4},\frac{3\pi}{4},\frac{5\pi}{4},\frac{7\pi}{4}$. Interestingly, in the two latter cases the phase transition at $\theta=\pi$ is absent, contrary to the pure YM case \cite{Gaiotto:2017yup}. Nevertheless, for the cases $N_F=0-2$ we show that all of our phase transitions can be explained by a slightly updated mixed-anomaly argument. {Besides the explicit expressions for the $\theta$-dependent potentials, our study also answers to the question of what dynamics is responsible for the generation of these somewhat unusual potentials with different branches and phase transitions between them. As foreseen by Witten, indeed they are not due to instantons, but rather by the condensation of various monopoles and dyons, which is also the mechanism of confinement itself. The origin of the branches lies in the existence of various monopoles/dyons, which will each have a $\theta$-dependent potential. As $\theta$ changes, the global minimum of these potentials moves from one set of vacua to the other, giving rise to first order phase transitions and branched structure of the QCD potential.}

The case of $N_F=3$ is unique in our work for several reasons. First, to the best of our knowledge we are the first to find the explicit form of the section $(a^{(3,2)}_D,a^{(3,2)})$ for which $a^{(3,2)}(z=0)=0$, and correspondingly the prepotential $F^{(3,2)}$ relevant for the monopole singularity near the origin. Our calculation is also the first to consider SUSY breaking and calculate the scalar potential for $N_F=3$. Thirdly, in the $N_F=3$ case the magnetic $U(1)_D$ ``magnetic" gauge symmetry is higgsed by a dyon of magnetic charge 2, leading to a $Z_2$ gauge theory in the IR. Finally, the $N_F=3$ case is unique in that it exhibits a phase transition from a chiral symmetry breaking phase (due to the condensation of a monopole in the $\mathbf{4}\in SU(4)$) to a chiral symmetry preserving TQFT (where the $(n_m,n_e)=(2,1)$ in the $\mathbf{1}\in SU(4)$) condenses.

The paper is structured as follows. In section~\ref{sec:summary} we summarize the main results of the paper, including the main plots of the vacuum energies as a function of $\theta$. In section~\ref{sec:SW} we provide a review of Seiberg-Witten (SW) theory for $N_F=0-3$, including known explicit results for the monodromies, prepotentials, and K\"ahler potentials. In section~\ref{sec:AMSB} we calculate the scalar potential of near-SUSY SW theory, from which we obtain the vacuum energy as a function of $\theta$. Finally, section~\ref{sec:comp} reviews the previous literature on softly broken $\mathcal{N}=2$ theories and their similarities and differences with the current work.

\section{Summary of Results}\label{sec:summary}
\subsection{Theta Dependent Vacuum Structure}\label{sec:thetavac}
\begin{figure}[h]
     \centering
     \begin{subfigure}[b]{0.45\textwidth}
         \centering
         \includegraphics[width=\textwidth]{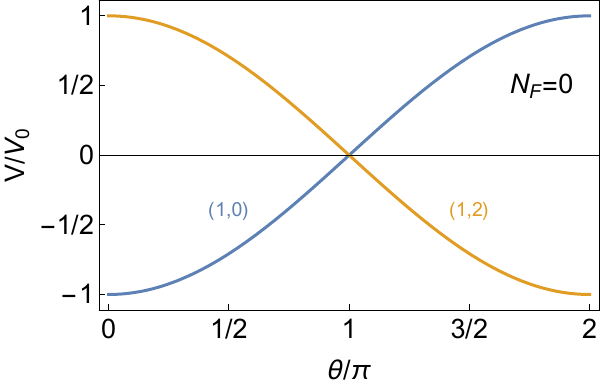}
         \caption{$N_F=0$}
         \label{fig:sNF0}
     \end{subfigure}
     \hfill
     \begin{subfigure}[b]{0.45\textwidth}
         \centering
         \includegraphics[width=\textwidth]{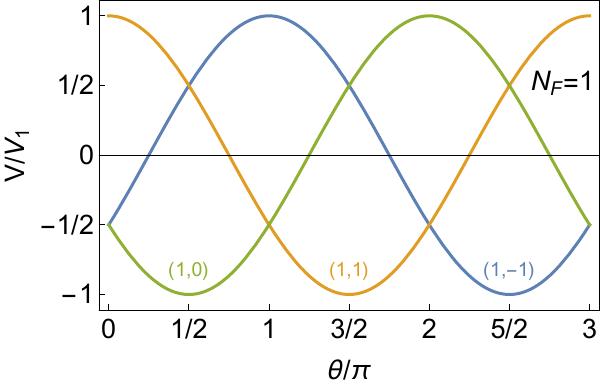}
         \caption{$N_F=1$}
         \label{fig:sNF1}
     \end{subfigure}
     \\ \vspace{10pt}
     \begin{subfigure}[b]{0.45\textwidth}
         \centering
         \includegraphics[width=\textwidth]{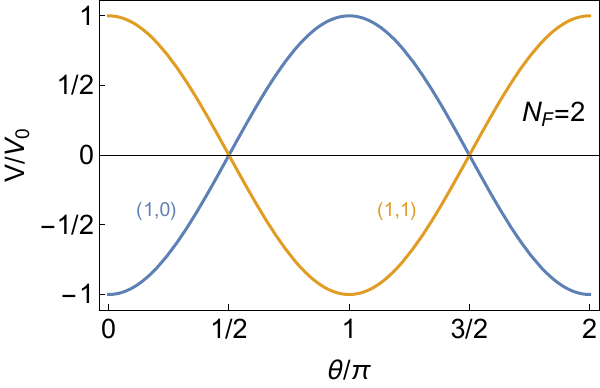}
         \caption{$N_F=2$}
         \label{fig:sNF2}
     \end{subfigure}
      \hfill
         \begin{subfigure}[b]{0.45\textwidth}
         \centering
         \includegraphics[width=\textwidth]{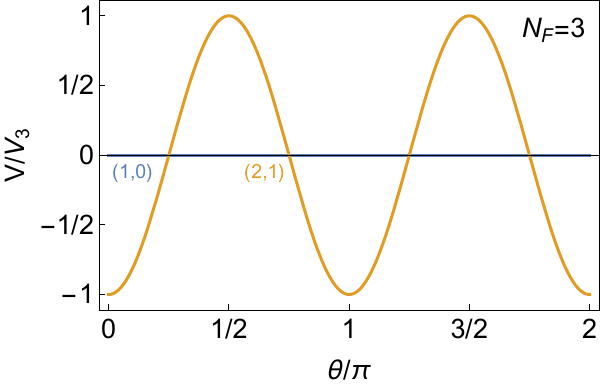}
         \caption{$N_F=3$}
         \label{fig:sNF3}
     \end{subfigure}
        \caption{Vacuum energies for $\mathcal{N}=2$ $SU(2)$ with a deformation to $\mathcal{N}=1$ and AMSB. The labels in blue, yellow and green are the charges $(n_m,n_e)$ of the monopoles which condense in each vacuum. Here $(V_0,V_1,V_2,V_3)=6m\mu\,(|\Lambda|^2,|\Lambda_1|^2,|\Lambda_2|^2,|\Lambda_3|^2)$.}
        \label{fig:NF0to3}
\end{figure}
The main results of the paper are depicted in Figure~\ref{fig:NF0to3}. Different colors represent different vacua in which different dyons condense. The magnetic and electric charges of the condensing dyons $(n_m,n_e)$ are depicted near each colored line. 

\subsection{$2\pi$ Periodicity in $\theta$ and the Witten Effect}\label{sec:thetaper}
\begin{figure}[h]
     \centering
     \begin{subfigure}[b]{0.45\textwidth}
         \centering
         \includegraphics[width=\textwidth]{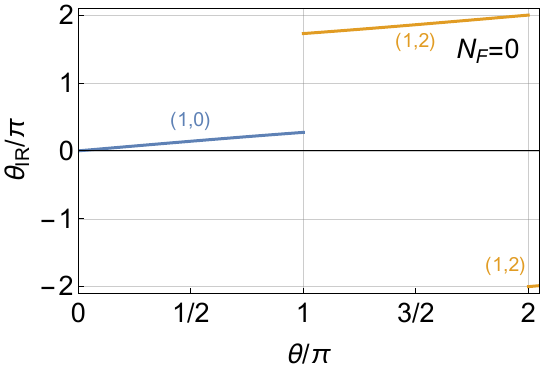}
         \caption{$N_F=0$}
         \label{fig:sNF0th}
     \end{subfigure}
     \hfill
     \begin{subfigure}[b]{0.45\textwidth}
         \centering
         \includegraphics[width=\textwidth]{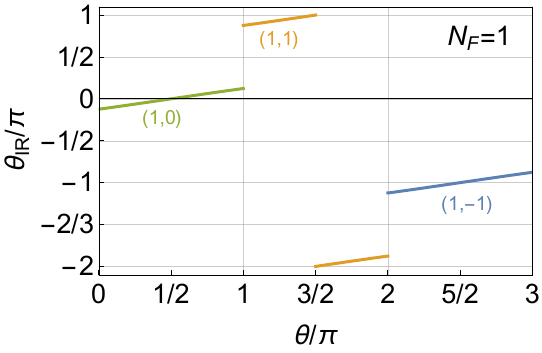}
         \caption{$N_F=1$}
         \label{fig:sNF1th}
     \end{subfigure}
     \\ \vspace{10pt}
     \begin{subfigure}[b]{0.45\textwidth}
         \centering
         \includegraphics[width=\textwidth]{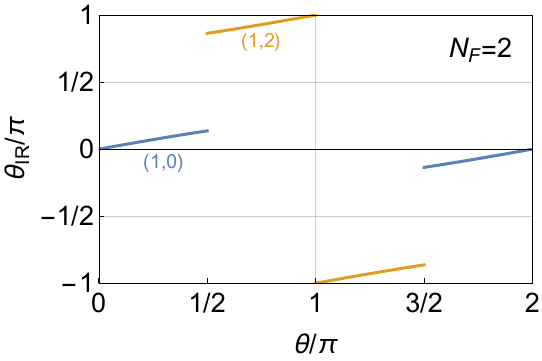}
         \caption{$N_F=2$}
         \label{fig:sNF2th}
     \end{subfigure}
      \hfill
         \begin{subfigure}[b]{0.45\textwidth}
         \centering
         \includegraphics[width=\textwidth]{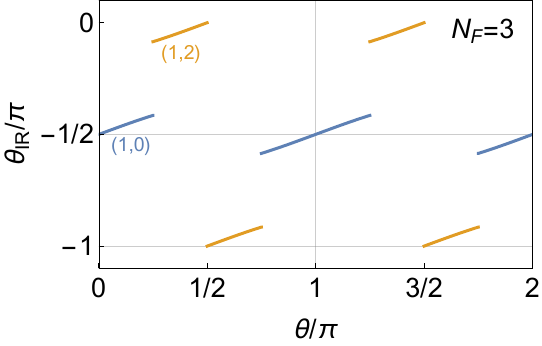}
         \caption{$N_F=3$}
         \label{fig:sNF3th}
     \end{subfigure}
        \caption{IR theta angle $\theta_{IR}=\pi\,{\rm Re}(\tau_{IR})$ for $\mathcal{N}=2$ $SU(2)$ with a deformation to $\mathcal{N}=1$ and AMSB. The labels in blue, yellow and green are the charges $(n_m,n_e)$ of the monopoles which condense in each vacuum. }
        \label{fig:NF0to3th}
\end{figure}

We explicitly checked that all of our physics result are invariant under $\theta\rightarrow\theta+2\pi$. The $2\pi$ periodicity is satisfied in a nontrivial way due to the Witten effect; the physical charge of the condensing dyon is given in these conventions by
\begin{eqnarray}
        (Q^{(N_F,k)}_m,Q^{(N_F,k)}_e)=(n^{(N_F,k)}_m,n^{(N_F,k)}_e-\frac{\theta^{(N_F,k)}_{IR}}{\pi}n^{(N_F,k)}_m)\,.
\end{eqnarray}
Here $\theta^{(N_F,k)}_{IR}=\pi\,{\rm Re}\left(\tau^{(N_F,k)}\right)$ is the IR theta angle at the $k$th vacuum\footnote{The normalization of $\tau$ and the Witten effect is in the conventions of \cite{Seiberg1994}.}. We explicitly calculated it below for all of our vacua, with the results shown in Figure~\ref{fig:NF0to3th}. These results, together with those of Figure~\ref{fig:NF0to3}, guarantee the $2\pi$ periodicity in $\theta$. Specifically, for $N_F=2\,(N_F=3)$, the $k=1\,(k=2)$ vacuum is the global minimum both at $\theta=0$ and $\theta=2\pi$. Accordingly, the value of $\theta_{IR}$ is the same in the $k=1\,(k=2)$ vacuum between $\theta=0$ and $\theta=2\pi$. Conversely, for $N_F=0$ the theory is in the $k=1$ vacuum with $(1,0)$ condensation for $\theta=0$, and in the $k=2$ vacuum with $(1,2)$ condensation for $\theta=2\pi$. However, due to the Witten effect, in both the $\theta=\varepsilon$ and $\theta=2\pi+\varepsilon$ (for an infinitesimal $\varepsilon>0$) the physical charge of the condensing dyons is $(Q_m,Q_e)=(1,0)$. Finally, for $N_F=1$ the theory is in the $k=3$ vacuum with $(1,0)$ condensation for $\theta=0$, and in the $k=1$ vacuum with $(1,-1)$ condensation for $\theta=2\pi$. Here as well, the Witten effect restores the $2\pi$ periodicity in $\theta$; in both the $\theta=\varepsilon$ and $\theta=2\pi+\varepsilon$, the physical charge of the condensing dyons is $(Q_m,Q_e)=\left(1,-\frac{\theta_{\rm IR}(\theta=0)}{\pi}\right)$.

\section{$\mathcal{N}=2$ Supersymmetric $SU(2)$ Gauge Theory}\label{sec:SW}
\subsection{General Features}
Consider $\mathcal{N}=2$ $SU(2)$ with $N_F$ matter fields $(Q_i,\widetilde{Q}_i)$ in the fundamental of the gauge group. $\mathcal{N}=2$ Supersymmetry constrains the superpotential to be of the form
\begin{eqnarray}
W=\sqrt{2}\,\widetilde{Q}_i\Phi Q^i\,,
\end{eqnarray}
where $\Phi$ is the chiral superfield part of the $\mathcal{N}=2$ gauge multiplet, in the adjoint of the gauge group $SU(2)$. 
This theory always has a Coulomb branch on which $\Phi$ gets a VEV and breaks $SU(2)$ to $U(1)$. There is no Higgs branch for $SU(2)$ with massless flavors. The Coulomb branch is parametrized by a coordinate $u$, which is related at weak coupling to the adjoint scalar $\phi$ in $\Phi$ as
\begin{eqnarray}
u=\left\langle\text{Tr}\phi^2\right\rangle~~~~~(\text{weak coupling}).
\end{eqnarray}
The K\"ahler potential is given on the Coulomb branch by:
\begin{eqnarray}\label{eq:KahN2}
K=\frac{1}{4\pi}\text{Im}\,a_D(u)\,\bar{a}(\bar{u})\,,
\end{eqnarray}
where $a_D(u),\,a(u)$ are sections of a holomorphic $SL(2,\mathbb{Z})$ bundle over the punctured $u$ complex plane. At weak coupling, $a$ is the photino while $a_D$ is the magnetic photino. The effective electric $U(1)$ gauge coupling is then given by
\begin{eqnarray}
\tau(a)=\partial^2_a\,F(a)~~,~~a_D=\partial_a F(a),\,
\end{eqnarray}
where $F(a)$ is the \textit{prepotential}. Equivalently, the \textit{magnetic} $U(1)$ gauge coupling is given by
\begin{eqnarray}
\tau_D(a)=\partial^2_{a_D}\,F_D(a_D)~~,~~a=-\partial_{a_D} F_D(a_D)\,.
\end{eqnarray}
Here $F_D$ is the \textit{magnetic prepotential}, and is equal to minus the Legendre transform of $F(a)$. Either $\tau$ or $\tau_D$ or both are singular at $u=\infty$ and at select points $u^{(N_F,k)}$ on the Coulomb branch - all of which are at strong coupling. These singularities signal the appearance of new massless states in the spectrum, whose number and electric and magnetic $U(1)$ charges are set by the monodromy of $\left(a_D(u),\,a(u)\right)$ around $u=\infty,u^{(N_F,k)}$. For example, when $N_F=0$, $\tau(a)$ in the weak duality frame is singular at $u=u^{(0,1)}\equiv\Lambda^2$, where $\Lambda$ is the strong scale of the theory. The corresponding monodromy is \cite{Seiberg1994,Seiberg1994a}
\begin{eqnarray}\label{eq:mon01}
&&\colvec{a_D(e^{2i\pi}u)\\a(e^{2i\pi}u)}=\mathcal{M}^0_1\,\colvec{a_D(u)\\a(u)},~~~~~~\mathcal{M}^0_1=STS^{-1}=\colmatt{0&1\\-1&1},\,
\end{eqnarray}
where
\begin{eqnarray}\label{eq:TS}
T\equiv\colmatt{1&1\\0&1},~~S\equiv\colmatt{0&1\\-1&0},\,
\end{eqnarray}
are the generators of $SL(2,\mathbb{Z})$. Having $T$ as the monodromy matrix indicates that particles with charges $(n_m,n_e)=(0,1)$ become massless. An $S$ duality transformation takes $T\rightarrow S T S^{-1}$ and $(n_m,n_e)\rightarrow S(0,1)=(1,0)$, and so the monodromy \eqref{eq:mon01} indicates that a monopole with $(n_m,n_e)=(1,0)$ becomes massless at $u=u^0_1\equiv\Lambda^2$.
For $0\leq N_F\leq 3$, we can conveniently encode all of the monodromy data in the Seiberg-Witten (SW) curve of the theory, which is derived from symmetry and holomorphy considerations. For $N_F=0$, the SW curve is
\begin{eqnarray}\label{eq:SW0}
y^2=x^2(x-u)+\frac{\Lambda^4}{4\,}x\,,
\end{eqnarray}
while for $N_F=1,2,3$ it is
\begin{eqnarray}\label{eq:SWNF}
y^2=x^2(x-u)-\frac{\Lambda^{2(4-N_F)}}{64}(x-u)^{N_F-1}\,.
\end{eqnarray}
Here $\Lambda=|\Lambda|e^{i\frac{\theta}{4-N_F}}$, where $4-N_F$ is the usual instanton factor. The SW curve encodes all of the IR structure of the $\mathcal{N}=2,\,SU(2)$ theory (see also \cite{Aspman:2021vhs} for a recent study of the modular structure of the theory).  The singularities on the Coulomb branch are exactly the values $u=u^{(N_F,k)}$ for which two or more roots $x_{a_1},\ldots,x_{a_n}$ of the SW curve coincide.\footnote{For $SU(N)$ theories the singularities always occur on a complex codimension one manifold, which could intersect~\cite{Argyres1995}.}  For finite $u^{(N_F,k)}$, the number of coinciding roots is always two. The section $\left(a_D,a\right)$ is then given by the periods of the SW curve: 
\begin{eqnarray}\label{eq:SWcyc}
a=\int_{\alpha}\,\lambda_{\text{SW}},~~a_D=\int_{\beta}\,\lambda_{\text{SW}},\,
\end{eqnarray}
where $\lambda_{\text{SW}}$ is the Seiberg-Witten differential
\begin{eqnarray}\label{eq:SWdif}
\lambda_{\text{SW}}=\frac{\sqrt{2}}{8\pi}\frac{2u-(4-N_F)x}{y}\,dx\,,
\end{eqnarray}
and $\alpha,\,\beta$ are curves in $x$, where each curve encloses a distinct pair of roots of the SW curve. $\alpha,\,\beta$ then form a basis for the space of homology cycles of the SW curve \eqref{eq:SW0}-\eqref{eq:SWNF}. A different choice of canonical curves leads to an $SL(2,\mathbb{Z})$ transformation of $\left(a_D,a\right)$.
\subsection{Singularity Structure}~\label{sec:sing}
Here we summarize the singularity structure of the moduli space for $0\leq N_F\leq 3$, as can be directly calculated from the SW curve.
They are expressed in terms of the $SL(2,\mathbb{Z})$ elements $T,\,S$ defined in \eqref{eq:TS}, as well as the inversion $P=-I_2$. The electric and magnetic charges $(n_m,n_e)$ are also given. 
For each $N_F$ we find sections $\left(a^{(N_F,k)}_D,a^{(N_F,k)}\right)$ so that
\begin{eqnarray}
&&\lim\limits_{u\rightarrow\infty}a_D^{(N_F,k)}\rightarrow\frac{1}{2}\sqrt{2u}\nonumber\\[5pt]
&&a^{(N_F,k)}(u^{(N_F,k)})=0\,.
\end{eqnarray}
i.e. $a^{(N_F,k)}$ is the local photon near the singularities $u^{(N_F,k)}$ while $a_D$ is the local photon at infinity.\footnote{Except for $N_F=3,k=1$ case, in which $a^{(3,1)}_D$ cannot be interpreted the photon at weak coupling. The  photon at weak coupling is then related to $(a^{(3,1)}_D,a^{(3,1)})$ by a duality transformation.} These can be obtained by explicitly evaluating the cycles \eqref{eq:SWcyc} \cite{AlvarezGaume1998}, or alternatively, by solving the Picard-Fuchs equations corresponding to the SW curve \eqref{eq:SW0}-\eqref{eq:SWNF}, subject to the monodromy conditions \cite{Klemm1995,Ito1996,Ito1996a,Bilal1996,Ferrari1996}. In appendix~\ref{app:derivation} we derive $\left(a^{(N_F,k)}_D,a^{(N_F,k)}\right)$ and $F^{(N_F,k)}$ using the latter method, following \cite{Klemm1995,Ito1996,Ito1996a,Bilal1996,Ferrari1996}, and in particular the solutions in particular \cite{Ferrari1996}. Our singularity structure is summarized below in tables for $0\leq N_F\leq 3$.
\begin{itemize}
\item $N_F=0$:
\begin{table}[h]
\centering
\begin{tabular}{|c|c|c|}
\hline
Singularity    & Monodromy            & $(n_m,n_e)$ \\ \hline\hline
$u=\infty$     & $PT^{4}$            & $(0,1)$\\ \hline
$u=u^{(0,1)}=~~\Lambda^2_0$ & $STS^{-1}$ &$(1,0)$\\ \hline
$u=u^{(0,2)}=-\Lambda^2_0$  & $(T^2S)T(T^2S)^{-1}$            & $(1,2)$\\ \hline
\end{tabular}\caption{$N_F=0$ singularities and their monodromies in the weak coupling frame.}\label{tab:NF0}
\end{table}
\vspace{5pt}\\
Note the $\mathbb{Z}_2$ symmetry exchanging the two strong coupling singularities. The photons at the singularities are given by
\begin{eqnarray}
&&a^{(0,1)}=-i\frac{u-\Lambda^2_0}{2\Lambda_0}+\mathcal{O}\left[(u-\Lambda^2)^2\right],\nonumber\\[5pt]
&&a^{(0,2)}=\frac{u+\Lambda^2_0}{2\Lambda_0}+\mathcal{O}\left[(u+\Lambda^2)^2\right]\,,
\end{eqnarray}
where $\Lambda_0=\Lambda$, while the prepotentials are given by
\begin{eqnarray}
&&F^{(0,1)}=-i\frac{a^{(0,1)}}{4\pi}\left[8i\Lambda_0+a^{(0,1)}\log\left(\frac{a^{(0,1)}}{\Lambda_0}\right)\right],\nonumber\\[5pt]
&&F^{(0,2)}=-i\frac{a^{(0,2)}}{4\pi}\left[8\Lambda_0+a^{(0,2)}\log\left(\frac{a^{(0,2)}}{\Lambda_0}\right)\right]\,.
\end{eqnarray}
\item $N_F=1$:
\begin{table}[h]
\centering
\begin{tabular}{|c|c|c|c|}
\hline
Singularity    & Monodromy            & $(n_m,n_e)$\\ \hline\hline
$u=\infty$     & $PT^{3}$            & $(0,1)$\\ \hline
$u=u^{(1,1)}=e^{i\pi/3}\Lambda^2_1$ & $(T^{-1}S)T(T^{-1}S)^{-1}$ & $(1,-1)$\\ \hline
$u=u^{(1,2)}=~-\,\Lambda^2_1$  & $(TS)T(TS)^{-1}$           & $(1,1)$\\ \hline
$u=u^{(1,3)}=e^{-i\pi/3}\Lambda^2_1$  & $STS^{-1}$           & $(1,0)$ \\ \hline
\end{tabular}\caption{$N_F=1$ singularities and their monodromies in the weak coupling frame.}\label{tab:NF1}
\end{table}
\vspace{5pt}\\
Here $\Lambda_1=\sqrt{3}\times2^{-4/3} \Lambda$. Note the $\mathbb{Z}_3$ symmetry exchanging the three strong coupling singularities. The photons at the singularities are given by
\begin{eqnarray}
&&a^{(1,1)}=e^{-2i\pi/3}\frac{u-u^{(1,1)}}{\sqrt{8}\Lambda_1}+\mathcal{O}\left[(u-u^{(1,1)})^2\right],\nonumber\\[5pt]
&&a^{(1,2)}=~-\,\frac{u-u^{(1,2)}}{\sqrt{8}\Lambda_1}+\mathcal{O}\left[(u-u^{(1,2)})^2\right],\nonumber\\[5pt]
&&a^{(1,3)}=-e^{2i\pi/3}\frac{u-u^{(1,3)}}{\sqrt{8}\Lambda_1}+\mathcal{O}\left[(u-u^{(1,3)})^2\right]\,.
\end{eqnarray}
The prepotentials are given by
\begin{eqnarray}
&&F^{(1,1)}=i\frac{a^{(1,1)}}{4\pi}\left[3\sqrt{8}e^{-i\pi/3}\Lambda_1-a^{(1,1)}\log\left(\frac{a^{(1,1)}}{\Lambda}\right)\right],\nonumber\\[5pt]
&&F^{(1,2)}=i\frac{a^{(1,2)}}{4\pi}\left[-3\sqrt{8}\Lambda_1-a^{(1,2)}\log\left(\frac{a^{(1,2)}}{\Lambda_1}\right)\right],\nonumber\\[5pt]
&&F^{(1,3)}=i\frac{a^{(1,3)}}{4\pi}\left[-3\sqrt{8}e^{i\pi/3}\Lambda_1-a^{(1,3)}\log\left(\frac{a^{(1,3)}}{\Lambda_1}\right)\right]\,.
\end{eqnarray}
\item $N_F=2$:
\begin{table}[h]
\centering
\begin{tabular}{|c|c|c|c|}
\hline
Singularity    & Monodromy            & Charge  &$SO(4)$\\ \hline\hline
$u=\infty$     & $PT^{-2}$            & $(0,1)$&-\\ \hline
$u=u^{(2,1)}=~\Lambda^2_{2}$ & $ST^2S^{-1}$           & $(1,0)$ &$(\mathbf{2},1)$\\ \hline
$u=u^{(2,2)}=-\Lambda^2_{2}$  & $(TS)T^2(TS)^{-1}$ & $(1,1)$&$(1,\mathbf{2})$ \\ \hline

\end{tabular}\caption{$N_F=2$ singularities and their monodromies in the weak coupling frame.}\label{tab:NF2}
\end{table}
\vspace{5pt}\\
Here $\Lambda_2=2^{-3/2}\Lambda$. Note again the $\mathbb{Z}_2$ symmetry exchanging the two strong coupling singularities, similarly to the $N_F=0$ case. The photons at the singularities are given by
\begin{eqnarray}
&&a^{(2,1)}=-i\frac{u-\Lambda^2_2}{4\Lambda_2}+\mathcal{O}\left[(u-\Lambda^2_2)^2\right],\nonumber\\[5pt]
&&a^{(2,2)}=\frac{u+\Lambda^2_2}{4\Lambda_2}+\mathcal{O}\left[(u+\Lambda^2_2)^2\right]\,,
\end{eqnarray}
while the prepotentials are
\begin{eqnarray}
&&F^{(2,1)}=-i\frac{a^{(2,1)}}{2\pi}\left[4i\Lambda_2+a^{(2,1)}\log\left(\frac{a^{(2,1)}}{\Lambda_2}\right)\right],\nonumber\\[5pt]
&&F^{(2,2)}=-i\frac{a^{(2,2)}}{2\pi}\left[4\Lambda_2+a^{(2,2)}\log\left(\frac{a^{(2,2)}}{\Lambda_2}\right)\right]\,.
\end{eqnarray}
\item $N_F=3$:
\begin{table}[h]
\centering
\begin{tabular}{|c|c|c|c|}
\hline
Singularity    & Monodromy            & Charge  &$SU(4)$\\ \hline\hline
$u=\infty$     & $PT$            & $(0,1)$&$-$\\ \hline
$u=u^{(3,1)}=0$  & $ST^4S^{-1}$           & $(1,0)$ &$\mathbf{4}$\\ \hline
$u=u^{(3,2)}=\Lambda^2_{3}$ & $(ST^2S)T(ST^2S)^{-1}$ & $(2,-1)$&$\mathbf{1}$\\ \hline
\end{tabular}\caption{$N_F=3$ singularities and their monodromies in the weak coupling frame.}\label{tab:NF3}
\end{table}
\vspace{5pt}\\
Here $\Lambda_3=2^{-4}\Lambda$. Note that unlike the $N_F=0,1,2$ cases, here there is no global symmetry relating the two singularities. The photons and prepotentials are given by
\begin{eqnarray}
&&a^{(3,1)}=\Lambda_3\frac{u}{2^{5/2}\Lambda^2_3}+\mathcal{O}\left[u^2\right],\nonumber\\[5pt]
&&a^{(3,2)}=i\Lambda_3\frac{u-u^3_2}{2^{3/2}\Lambda^2_3}+\mathcal{O}\left[(u-u^3_2)^2\right],\nonumber\\[5pt]
&&F^{(3,1)}=i\frac{a^{(3,1)}}{\pi}\left[-\sqrt{2}\Lambda_3-a^{(3,1)}\log\left(\frac{a^{(3,1)}}{\Lambda_3}\right)\right],\nonumber\\[5pt]
&&F^{(3,2)}=i\frac{a^{(3,2)}}{8\pi}\left[i\sqrt{8}\Lambda_3-a^{(3,2)}\log\left(\frac{a^{(3,2)}}{\Lambda_3}\right)\right]\,.
\end{eqnarray}
\end{itemize}
\vspace{5pt}
Once the section $\left(a^{(N_F,k)}_D(u),a^{(N_F,k)}(u)\right)$ is found, the moduli space of the $\mathcal{N}=2$ theory is practically solved. In particular, the K\"ahler potential on the Coulomb branch is given by \eqref{eq:KahN2}, while the effective superpotential at each strong coupling singularity is 
\begin{eqnarray}\label{eq:sp}
W_{u^{(N_F,k)}}=\sqrt{2}\,a^{(N_F,k)}\,M_n\,\widetilde{M}_n\,.
\end{eqnarray}
here for $M_n\,\tilde{M}_n$ are the two~~$\mathcal{N}=1$ chiral superfields that make up a $\mathcal{N}=2$ hypermultiplet of dyons, that become massless at $u=u^{(N_F,k)}$. These come in $n$ flavors, where $n$ is the dimension of the dyon flavor representations. The flavor representations appear in the rightmost columns of the tables of the current section. 

\subsection{K\"ahler Potentials and Their Derivatives}
In the $u^{(N_F,k)}$ duality frame, the K\"ahler potential for $a^{(N_F,k)}$ is given by 
\begin{eqnarray}\label{eq:KD}
	K^{(N_F,k)}=\frac{1}{4\pi}\text{Im}\left(a^{(N_F,k)}\bar{a}^{(N_F,k)}_{D}\right).
\end{eqnarray}
Extracting $\frac{du}{da^{(N_F,k)}}$, $\frac{\partial K^{(N_F,k)}}{\partial{a^{(N_F,k)}}}$ and $g^2_{(N_F,k)}=\frac{8\pi}{{\rm Im} \,\tau^{(N_F,k)}}$ , and expanding them to leading order in $u-u_{\rm singularity}$, we get
\begin{eqnarray}\label{eq:KDNF0t}
	&&\frac{\partial K^{(0,1)}}{\partial{a^{(0,1)}}}=~\frac{i\bar{\Lambda}_0}{4\pi^2}~~,~~\frac{du}{da^{(0,1)}}=2i \Lambda_0,\nonumber\\[5pt]
 &&\frac{\partial K^{(0,2)}}{\partial{a^{(0,2)}}}=-\frac{\bar{\Lambda}_0}{4\pi^2}~~,~~\frac{du}{da^{(0,2)}}=2 \Lambda_0\,.
\end{eqnarray}
as well as
\begin{eqnarray}\label{eq:g0}
	g^2_{(0,k)}=\frac{32\pi^2}{\log{\left|\frac{\Lambda_0}{\Lambda^{(0,k)}_{IR}}\right|}^2-1}~~~~~~(k=1,2)\,.
\end{eqnarray}
Note that $\Lambda^{(0,k)}_{IR}\equiv|a^{(0,k)}(u)|$ vanishes at the monopole/dyon singularities on the moduli space. Nevertheless, we will see below that the actual vacua of the (SUSY broken) theory are slightly off the monopole point by a small amount $\mathcal{O}(m/\Lambda)$, and consequently $\Lambda^{(N_F,k)}_{IR}$ is stabilized at $m\ll \Lambda_{N_F}$.
Similarly to the $N_F=0$ case, in the $N_F=1$ we get
\begin{eqnarray}\label{eq:KDNF1t}
	&&\frac{\partial K^{(1,1)}}{\partial{a^{(1,1)}}}=e^{i\pi/3}\,\frac{3\sqrt{2}\bar{\Lambda}_1}{16\pi^2}~~~,~~\frac{du}{da^{(1,1)}}=\sqrt{8}e^{2i\pi/3}\Lambda_1,\nonumber\\[5pt]
 &&\frac{\partial K^{(1,2)}}{\partial{a^{(1,2)}}}=-\frac{3\sqrt{2}\bar{\Lambda}_1}{16\pi^2}~~~~~~~,~~\frac{du}{da^{(1,2)}}=-\sqrt{8}\Lambda_1,\nonumber\\[5pt]
 &&\frac{\partial K^{(1,3)}}{\partial{a^{(1,3)}}}=e^{-i\pi/3}\,\frac{3\sqrt{2}\bar{\Lambda}_1}{16\pi^2}~~,~~\frac{du}{da^{(1,3)}}=\sqrt{8}e^{-2i\pi/3}\Lambda_1\,,
 \end{eqnarray}
as well as
\begin{eqnarray}\label{eq:g1}
	g^2_{(1,k)}=\frac{32\pi^2}{\log{\left|\frac{\Lambda_1}{\Lambda^{(0,k)}_{IR}}\right|}^2-1}~~~~~~(k=1,2,3)\,.
\end{eqnarray}
The corresponding result for $N_F=2$ is
\begin{eqnarray}
    &&\frac{\partial K^{(2,1)}}{\partial{a^{(2,1)}}}=~i\frac{\bar{\Lambda}_2}{4\pi^2}~~~,~~~\frac{du}{da^{(2,1)}}=4i\Lambda_2,\nonumber\\[5pt]
    &&\frac{\partial K^{(2,2)}}{\partial{a^{(2,2)}}}=-\frac{\bar{\Lambda}_2}{4\pi^2}~~~,~~~\frac{du}{da^{(2,2)}}=4\Lambda_2\,,
    \end{eqnarray}
    as well as
\begin{eqnarray}\label{eq:g2}
	g^2_{(2,k)}=\frac{16\pi^2}{\log{\left|\frac{\Lambda_2}{\Lambda_{IR}}\right|}^2-1}~~~~~~(k=1,2)\,.
\end{eqnarray}
Finally for $N_F=3$ one finds
    \begin{eqnarray}
     &&\frac{\partial K^{(3,1)}}{\partial{a^{(3,1)}}}=-\frac{1}{\sqrt{2}}\frac{\bar{\Lambda}_3}{4\pi^2}~~,~~\frac{du}{da^{(3,1)}}=2\sqrt{8}\Lambda_3,\nonumber\\[5pt]
     &&\frac{\partial K^{(3,2)}}{\partial{a^{(3,2)}}}=-\frac{i}{\sqrt{8}}\frac{i\bar{\Lambda}_3}{4\pi^2}~~,~~\frac{du}{da^{(3,2)}}=-i\sqrt{8}\Lambda_3\,,
\end{eqnarray}
    as well as
\begin{eqnarray}\label{eq:g3}
	g^2_{(3,1)}=\frac{8\pi^2}{\log{\left|\frac{\Lambda_3}{\Lambda_{IR}}\right|}^2-1}\,,\nonumber\\[5pt]
 g^2_{(3,2)}=\frac{32\pi^2}{\log{\left|\frac{\Lambda_3}{\Lambda_{IR}}\right|}^2-1}\,.
\end{eqnarray}
\section{Deforming to $\mathcal{N}=1$ and adding AMSB}\label{sec:AMSB}

Before introducing AMSB, we first introduce a deformation of the theory to $\mathcal{N}=1$. This is easily achieved by introducing the gaugino mass term
\begin{eqnarray}\label{eq:DW}
\Delta W=\mu \,u\,.
\end{eqnarray}
In the large $\mu$ limit, this has the effect of decoupling the gaugino. In turn, the Coulomb branch is eliminated, and only the singularities $u=u^{(N_F,k)}$ remain as $\mathcal{N}=1$ supersymmetric minima \cite{Seiberg1994a,Seiberg1994}. In these minima, the dyons get a negative mass and condense. To see this let us write down the scalar potential around $u^{(N_F,k)}$. It is given by
\begin{eqnarray}
V_{u^{(N_F,k)}}&=&\,\frac{g^2_{(N_F,k)}}{2}\,\left(\left|\sqrt{2}M_n\widetilde{M}_n+\mu\frac{du}{da^{(N_F,k)}}\right|^2+(|M_n|^2-|\widetilde{M}_n|^2)^2\right)\, \nonumber \\
&+& 2|a^{(N_F,k)}|^2\,\left(|M_n|^2+|\widetilde{M}_n|^2\right)\, .
\end{eqnarray}
The $\mathcal{N}=1$ supersymmetric minimum is clearly at $a^{(N_F,k)}=0,\,M_n\widetilde{M}_n=-\frac{\mu}{\sqrt{2}}\frac{du}{da^{(N_F,k)}}$, and $|M_n|=|\widetilde{M}_n|$. In the strict $\mu\rightarrow\infty$ limit, the theory reduces to $\mathcal{N}=1$ with $N_F$ flavors, and a Higgs branch appears. In this case the vacuum structure of the theory in the near-SUSY limit can be studied by adding AMSB to the $\mathcal{N}=1$ theory \cite{Murayama:2021xfj,Csaki2022b}, and so we do not consider the strict $\mu\rightarrow\infty$ limit in this paper.

We now add AMSB to the theory by coupling it to the conformal compensator $\varepsilon=1+m\theta^2$. This leads to a new SUSY breaking contribution \cite{Pomarol1999} (see also \cite{Kondo:2021osz,Csaki:2021xuc}) to the scalar potential near $u^{(N_F,k)}$,
\begin{eqnarray}\label{eq:extendedAMSB}
V^{(N_F,k)}_{\rm tot}&=&V_{u^{(N_F,k)}}+\Delta V^{(N_F,k)}_{\rm AMSB},\nonumber\\[5pt]
	\Delta V_{\rm AMSB}&=&m \left(\partial_i W K^{i \bar{j}} \partial_{\bar{j}} K - 3 W \right) + c.c.\nonumber\\
	&&+ \,|m|^2 \left( \partial_i K K^{i \bar{j}} \partial_{\bar{j}} K - K \right)\,.
\end{eqnarray}
Here we use the standard shorthand notation $\partial_i\equiv\frac{\partial}{\partial\phi_i}$ where $\phi_i$ is the $i$-th scalar field.
$K_{i\bar{j}}\equiv\partial_i\partial_{\bar{j}}K$ is the K\"ahler metric and $K^{i\bar{j}}$ is the matrix inverse of $K_{i\bar{j}}$. Substituting the superpotential \eqref{eq:sp}-\eqref{eq:DW}, we get
\begin{eqnarray}\label{eq:extendedAMSB2}
	&&\Delta V^{(N_F,k)}_{\rm AMSB}=\frac{g^2_{(N_F,k)}}{2}\,\left[m \,\frac{\partial K^{(N_F,k)}}{\partial{\bar{a}^{(N_F,k)}}}\left(\sqrt{2}M_n\widetilde{M}_n+\mu\frac{du}{da^{(N_F,k)}}\right) + c.c. \right.
 \nonumber \\ &+& \left. |m|^2 \frac{\partial K^{(N_F,k)}}{\partial{a^{(N_F,k)}}}\frac{\partial K^{(N_F,k)}}{\partial{\bar{a}^{(N_F,k)}}}\right]
 -|m|^2K^{(N_F,k)}-m\left[\sqrt{2}\,a^{(N_F,k)}\,M_n\widetilde{M}_n+3\mu u\right]+\text{cc.}\,.\nonumber \\
\end{eqnarray}

From this general form we can already learn that the minima will be at values of $u$ which are offset from the monopole points by $\mathcal{O}(m/\Lambda_{N_F})\times\Lambda^2_{N_F}$, namely $u^{(N_F,k)}_{min}=u^{(N_F,k)}+A^{(N_F,k)}m\Lambda_{N_F}$ for $A^{(N_F,k)}=\mathcal{O}(1)$. For the purpose of finding the monopole condensate and vacuum energy, this small deviation from the dyon singularities only contributes at higher orders and can be neglected. Only the sign of the deviation $A^{(N_F,k)}$ is important to calculate the IR theta angle $\theta^{(N_F,k)}_{IR}=\pi {\rm Re}\,{\tau^{(N_F,k)}}$ at $u^{(N_F,k)}_{min}$.

\subsection{Non-Supersymmetric Minima}

Substituting the explicit expression from the last section in \eqref{eq:extendedAMSB}, we get the scalar potentials for different $N_F$. For notational compactness, we project this potential on the $D$-flat direction where $|M^{(N_F,k)}_n|=|\widetilde{M}^{(N_F,k)}_n|$. Since we are only interested in leading order values in $m/\Lambda$ for the condensates, we minimize the potential using the following self-consistent algorithm: 
\begin{enumerate}
    \item Set $u=u^{(N_F,k)}+A^{(N_F,k)}m\Lambda_{N_F}$ for the still unknown value $A^{(N_F,k)}$ of the deviation form the dyon singularities.
    \item Find the condensate $M^{(N_F,k)}_n$ and vacuum energy. To leading order, they are independent of $A^{(N_F,k)}$.
    \item Substitute $M^{(N_F,k)}_n$ and find $A^{(N_F,k)}$.
\end{enumerate}
We now show how this algorithm works explicitly for all $(N_F,k)$.
\begin{itemize}
\item $N_F=0$:\\\quad\\
Setting $u=u^{(0,k)}+A^{(0,k)}m\Lambda_{0}$, we have
\begin{eqnarray}
&&V^{(0,k)}_{\text{tot}}=g^2_{(0,k)}\left|(M^{(0,k)})^2-\sqrt{2}(-i)^k\left(\mu\Lambda_0-(-1)^k\frac{\bar{m}\bar{\Lambda}_0}{8\pi^2}\right)\right|^2+\mathcal{O}\left(m\mu^2\Lambda_0\right)\nonumber\\[5pt]
	&&~~~~~~~~~+(-1)^k\,3\left[m\mu \Lambda^2_0+cc.\right]\,.
\end{eqnarray}
The minima are then at
\begin{eqnarray}\label{eq:mon0}
&&(M^{(0,k)})^2=\sqrt{2}(-i)^k\left(\mu\Lambda_0-(-1)^k\frac{\bar{m}\bar{\Lambda}_0}{8\pi^2}\right)+\mathcal{O}\left(m\mu\right)\,.
\end{eqnarray}
Without loss of generality, we can take $\mu,\,m$ to be real, while the complex argument of $\Lambda_0$ is $\Lambda_0=|\Lambda_0|e^{i\frac{\theta}{4-N_F}}$, where $4-N_F$ is the usual instanton factor. The vacuum energy at the minima for $k=1,2$ is then given by
\begin{eqnarray}\label{eq:N2NF0vac}
V^{(0,k)}_{\text{min}}=-6m\mu|\Lambda_0|^2 \cos\left(\frac{\theta}{2}+(k-1)\pi\right)+\mathcal{O}\left(m\mu^2\Lambda_0\right)\,.
\end{eqnarray}
Clearly, the $k=1$ minimum is the global one for $0<\theta<\pi$. In this vacuum the monopoples with charge $(n_m,n_e)=(1,0)$ condense. On the other hand,for $\pi<\theta<2\pi$ the global minimum is the $k=2$ one, in which the dyons $(n_m,n_e)=(1,2)$ condense. At $\theta=\pi$ there is a first order phase transition between the two vacua, a result first obtained in \cite{Konishi1997,Evans1997}. The vacuum energy as a function of $\theta$ is depicted in figure~\ref{fig:sNF0}. This result also qualitatively agrees with Witten's original prediction~\cite{Witten:1978bc,Witten:1980sp} for the vacuum structure of QCD without quarks, obtained based on large-$N$ considerations. Witten found that the potential is a function of $\theta /N$ with $N$ branches for $SU(N)$ Yang-Mills, and only becomes $2\pi$ periodic in $\theta$ after minimizing over the various branches. This also implies that the $\theta$ dependence is not exclusively due to instanton effects, because those would not result in a branched structure. Indeed we can see here that the potential is a consequence of monopole/dyon condensation, and Witten's qualitative picture is explicitly realized. 
Finally, we substitute the VEV \eqref{eq:mon0} in \eqref{eq:extendedAMSB}, and minimize it to get $A^{(0,k)}=\sqrt{2}$.
Evaluating $\theta^{(0,k)}_{IR}=\pi\,{\rm Re}\,\tau^{(0,k)}$ at the minimum $u=u^{(0,k)}+A^{(0,k)}m\Lambda_{0}$, we get the result presented in Figure~\eqref{fig:sNF0th}. This guarantees the $2\pi$ periodicity in $\theta$, as explained in section~\ref{sec:thetaper}.

\item $N_F=1$:\\\quad\\
Setting $u=u^{(1,k)}+A^{(1,k)}m\Lambda_{1}$, we have
\begin{eqnarray}
&&V^{(1,k)}_{\text{tot}}=g^2_{(1,k)}\left|(M^{(1,k)})^2+\frac{1}{2}\left(4e^{-2i(k-2)\pi/3}\mu\Lambda_1-e^{2i(k-2)\pi/3}\frac{3\bar{m}\bar{\Lambda}_1}{8\pi^2}\right)\right|^2+\mathcal{O}\left(m\mu^2\Lambda_1\right)\nonumber\\[5pt]
	&&~~~~~~~~~+3\,\left[ e^{-2i(k-2)\pi/3}m\mu\Lambda^2_1+\rm cc.\right]\,.
\end{eqnarray}
The minima are then at
\begin{eqnarray}\label{eq:minimaN2NF1}
&&(M^{(1,k)})^2=-\frac{1}{2}\left(4e^{-2i(k-2)\pi/3}\mu\Lambda_1-e^{2i(k-2)\pi/3}\frac{3\bar{m}\bar{\Lambda}_1}{8\pi^2}\right)+\mathcal{O}\left(m\mu\right)\,,
\end{eqnarray}
and the vacuum energy for $k=1,2,3$ is
\begin{eqnarray}\label{eq:N2NF1vac}
V^{(1,k)}_{\text{min}}=6m\mu|\Lambda_1|^2\cos \left(\frac{2\left[\theta+(k-2)\pi\right]}{3}\right)+\mathcal{O}\left(m\mu^2\Lambda_1\right)\,.
\end{eqnarray}
The vacuum energy for the different minima is presented in figure~\ref{fig:sNF1}. In particular, the system goes between the $k=3$ minimum with $(1,0)$ monopole condensation for $0<\theta<\pi$, the $k=2$ minimum with $(1,1)$ dyon condensation for $\pi<\theta<2\pi$, and the $k=1$ minimum with $(1,-1)$ dyon condensation for $2\pi<\theta<3\pi$. Finally, we substitute the VEV \eqref{eq:minimaN2NF1} in \eqref{eq:extendedAMSB}, and minimize it to get $A^{(1,k)}=2$. Evaluating $\theta^{(1,k)}_{IR}=\pi\,{\rm Re}\,\tau^{(1,k)}$ at the minimum $u=u^{(1,k)}+A^{(1,k)}m\Lambda_{1}$, we get the result presented in Figure~\eqref{fig:sNF1th}. This guarantees the $2\pi$ periodicity in $\theta$, as explained in section~\ref{sec:thetaper}.

\item $N_F=2$:\\\quad\\
Setting $u=u^{(2,k)}+A^{(2,k)}m\Lambda_{2}$, we have
\begin{eqnarray}
&&V^{(2,k)}_{\text{tot}}=g^{2~(2,k)}\left|(M^{(2,k)}_n)^2-\sqrt{8}\,(-i)^k\left(\mu\Lambda_2-(-1)^k\frac{1}{2}\frac{\bar{m}\bar{\Lambda}_2}{8\pi^2}\right)\right|^2+\mathcal{O}\left(m\mu^2\Lambda_2\right)\nonumber\\[5pt]
	&&~~~~~~~~~-(-1)^k\,3\left[m\mu \Lambda^2_2+cc.\right]\,.
\end{eqnarray}
Here $n=1,2$ is the monopole flavor index. The minima are then at
\begin{eqnarray}\label{eq:mon2}
&&(M^{(2,k)}_n)^2=\sqrt{8}\,(-i)^k\left(\mu\Lambda_2-(-1)^k\frac{1}{2}\frac{\bar{m}\bar{\Lambda}_2}{8\pi^2}\right)+\mathcal{O}\left(m\mu\right)\,.
\end{eqnarray}
The vacuum energy at the minima for $k=1,2$ is then given by
\begin{eqnarray}
V^{(2,k)}_{\text{min}}=-6m\mu|\Lambda_2|^2 \cos\left(\theta+(k-1)\pi\right)+\mathcal{O}\left(m^2\mu^2\right)\,.
\end{eqnarray}
The vacuum energy for the different minima is presented in figure~\ref{fig:sNF2}. In particular, the system goes between the $k=1$ minimum with $(1,0)$ monopole condensation for $0<\theta<\frac{\pi}{2}$ and $\frac{3\pi}{2}<\theta<2\pi$, the $k=2$ minimum with $(1,1)$ dyon condensation for $\frac{\pi}{2}<\theta<\frac{3\pi}{2}$. These minima involve the condensation of monopoles/dyons in the $(\mathbf{2},1)$ or $(1,\mathbf{2})$ flavor symmetry, and so they lead to chiral symmetry breaking $SU(2)_L\times SU(2)_R\rightarrow SU(2)_L$ or $SU(2)_R$. Substituting the VEV \eqref{eq:mon2} in \eqref{eq:extendedAMSB}, and minimizing, we get $A^{(2,k)}=\sqrt{8}$.
Once again evaluating $\theta^{(2,k)}_{IR}=\pi\,{\rm Re}\,\tau^{(2,k)}$ at the minimum $u=u^{(2,k)}+A^{(2,k)}m\Lambda_{2}$, we get the result presented in Figure~\eqref{fig:sNF2th}. This again guarantees the $2\pi$ periodicity in $\theta$.

\item $N_F=3$:\\\quad\\
For the singularity at the origin we set $u=A^{(3,1)}m\Lambda_3$, so that
\begin{eqnarray}
&&V^{(3,1)}_{\text{tot}}=g^{2~(3,1)}\left|(M^{(3,1)}_n)^2+\left(4\mu\Lambda_3-\frac{\bar{m}\bar{\Lambda}_3}{8\pi^2}\right)\right|^2+\mathcal{O}\left(m\mu^2\Lambda_2\right)\,,
\end{eqnarray}
while for the second singularity we set $u=u^{(3,2)}+{A^{(3,2)}} m\Lambda_{3}$, we have
\begin{eqnarray}
&&V^{(3,2)}_{\text{tot}}=g^{2~(3,2)}\left|{(M^{(3,2)})}^2-\frac{i}{2}\left(4\mu\Lambda_3+\frac{\bar{m}\bar{\Lambda}_3}{8\pi^2}\right)\right|^2+\mathcal{O}\left(m\mu^2\Lambda_3\right)\nonumber\\[5pt]
 &&~~~~~~~~~-3\left[m\mu \Lambda^2_3+cc.\right]\,.
\end{eqnarray}
Here $n=1,\ldots,4$ is the flavor index of the monopoles at $u=0$. The minimum near the $u=0$ singularity is then
\begin{eqnarray}\label{eq:mon31}
&&(M^{(3,1)}_n)^2=-\left(4\mu\Lambda_3-\frac{\bar{m}\bar{\Lambda}_3}{8\pi^2}\right)+\mathcal{O}\left(m\mu\right)\,,
\end{eqnarray}
While the monopole VEV close to the $u=\Lambda^2_3$ singularity is
\begin{eqnarray}\label{eq:mon32}
&&(M^{(3,2)})^2=\frac{i}{2}\left(4\mu\Lambda_3-\frac{\bar{m}\bar{\Lambda}_3}{8\pi^2}\right)+\mathcal{O}\left(m\mu\right)\,.
\end{eqnarray}
The vacuum energy at these minima is then given by
\begin{eqnarray}
V^{(3,1)}_{\text{min}}=\mathcal{O}\left(m\mu^2\Lambda_3\right)~~~,~~~V^{(3,2)}_{\text{min}}=-6m\mu|\Lambda_3|^2 \cos\left(2\theta\right)+\mathcal{O}\left(m\mu^2\Lambda_3\right)\,.
\end{eqnarray}
\end{itemize}
The vacuum energy for the different minima is presented in figure~\ref{fig:NF0to3}. In particular, the system goes between the $k=1$ minimum with the condensing $(1,0)$ monopoles in the $\mathbf{4}$ of $SU(4)$, and the $k=2$ minimum with the condensing of $(2,1)$ monopoles in the $\mathbf{1}$ of $SU(4)$. The $k=1$ minimum is the global one for $\frac{\pi}{4}<\theta<\frac{3\pi}{4}$ and $\frac{5\pi}{4}<\theta<\frac{7\pi}{4}$, while the $k=2$ minimum is the global one for $0<\theta<\frac{\pi}{4}$, $\frac{3\pi}{4}<\theta<\frac{5\pi}{4}$, and $\frac{7\pi}{4}<\theta<2\pi$. Once again, we substitute the VEVs \eqref{eq:mon31} and \eqref{eq:mon32} in \eqref{eq:extendedAMSB}, and minimize it to get $A^{(3,1)}=4$ and $A^{(3,2)}=2$.
Evaluating $\theta^{(3,k)}_{IR}=\pi\,{\rm Re}\,\tau^{(3,k)}$ at the minimum $u=u^{(3,k)}+A^{(3,k)}m\Lambda_{3}$, we get the result presented in Figure~\eqref{fig:sNF3th}. This guarantees the $2\pi$ periodicity in $\theta$ for $N_F=3$.

\section{Comparisons With Literature}\label{sec:comp}
In this section, we comment on the similarities and differences between the current work and previous studies of softly broken $\mathcal{N}=2$ gauge theory. 
\begin{itemize}
    \item The pioneering works are \cite{Konishi1997} and \cite{Evans1997}. These authors considered $\mathcal{N}=2$ $SU(2)$ gauge theory with $\mathcal{N}_F=0$ flavors and the $\mathcal{N}=1$ deformation $\mu u$. In addition, they added ``by hand" a soft mass for the scalar components of the monopoles/dyons at $u=\pm\Lambda^2$. As a result, they got the vacuum structure depicted in Figure~\ref{fig:NF0to3} for $N_F=0$, including the phase transition at $\theta=\pi$. Our results for $N_F=0$ reproduce their original results, with the main difference that our soft breaking originates from AMSB and is mapped exactly from the UV to the IR theory, rather than being put ``by hand".
    \item The series of works \cite{AlvarezGaume1996,AlvarezGaume1997,AlvarezGaume1998,AlvarezGaume1998a} considered $\mathcal{N}=2$ $SU(2)$ gauge theory coupled to holomorphic SUSY breaking spurions. These works did not introduce an explicit deformation to $\mathcal{N}=1$. In particular, reference \cite{AlvarezGaume1996} considered coupling the $\mathcal{N}=2$ theory to a dilaton via
    \begin{eqnarray}
    \Lambda\rightarrow \Lambda e^{iS}\,,
    \end{eqnarray}
    where $S$ is a vector superfield whose auxiliary field acts as a SUSY breaking spurion. Using holomorphy, the coupling to $S$ is easily mapped to the EFT near the monopole/dyon points, allowing to study their SUSY breaking vacua for $N_F=0,2$ \cite{AlvarezGaume1996} and for $N_F=1$ \cite{AlvarezGaume1997}. These works only considered the case in which the bare $\theta=0$. Accordingly, for $N_F=0,2$, the authors find a global minimum with monopole (rather than dyon) condensation, while for $N_F=1$ the vacuum has two degenerate minima near $u=e^{\pm i\pi/3}\Lambda^2$ where the $(n_m,n_e)=(1,0)$ and $(1,1)$ dyons condense. Finally, references \cite{AlvarezGaume1998,AlvarezGaume1998a} generalized the analysis to include $N_F=1,2$ massive flavors coupled to the dilaton-spurion $S$ and additional SUSY breaking from the $F$-terms of the bare masses $m$. The IR phase in this case involves monopole condensation, dyon condensation or quark condensation, depending on the bare masses and the soft terms. The main differences between these papers and the present work are the consideration of only holomorphic SUSY breaking spurions, the absence of explicit breaking to $\mathcal{N}=1$, and the fact that the bare $\theta$ parameter is taken to vanish in the former.

    \item The well-known work \cite{Luty1999} (see also the earlier \cite{ArkaniHamed1998}) presented two ways of exactly mapping non-holomorphic soft terms from the UV to the IR. The first way is based on the definition of two RG invariant spurions: a chiral superfield $\Lambda_S$ charged under an anomalous $U(1)$, and a real superfield $\Lambda_R$. There components of these two spurions are determined by the UV soft terms, while the way they enter the IR theory is fixed by dimensional analysis and the anomalous $U(1)$. This allows for the exact mapping of non-holomorphic soft terms to the IR. The second way is coupling the theory to a SUGRA background with a gauged $U(1)_R$. In this case, the soft terms in the UV can be mapped to the $D$-term of the gauge field $V_R$ and the $F$-term of the conformal compensator $\phi$. Since the SUGRA coupling is fixed both in the UV and the IR, this allows the mapping of the UV soft terms to the IR. Note that the AMSB method we are using in this paper is a special case of the latter method when only the compensator $F$-term is turned on. For $\mathcal{N}=2$ with $N_F=0$, \cite{Luty1999} used the first method based on $\Lambda_{S,R}$, and included all possible soft breaking terms. We checked that, indeed, our scalar potential for $N_F=0$ is a special case\footnote{Note that the solution in \cite{Luty1999} is given in terms of the `old' Seiberg-Witten conventions of \cite{Seiberg1994}, while ours are in the revised conventions of \cite{Seiberg1994}.} of their formalism when we set their soft parameters to be $g^{-2}_0\left(m_{\phi0},m_{\lambda0},m_{\chi0},m_{B0}\right)=\left(0,\frac{m}{4\pi^2},-\mu,\sqrt{\mu m}\right)$.

    \item The paper \cite{Abel2011} refined the exact mapping of soft parameters from the UV to the IR explored in \cite{Luty1999}. As in the latter paper, the mapping of non-holomorphic data was again carried out in two complementary ways. The first one is by embedding the soft terms in the bottom component of the Ferrara-Zumino (FZ) anomalous supermultiplet. The latter can be mapped to the IR theory because its divergence is a superfield containing the energy-momentum tensor. The second method, which generalizes \cite{Luty1999}, is valid when the theory has a conserved $R$-current, which is taken to be the bottom component of an $R$-supercurrent multiplet. In that case, we can embed the soft terms in the derivative of the $R$-supercurrent multiplet. The paper surveys several examples for the mapping of soft terms in $\mathcal{N}=1$ SUSY gauge theory, but does not explicitly work out $\mathcal{N}=2$.

    \item The work \cite{Cordova2018} is a modern and thorough study of $N_F=0$ Seiberg-Witten theory and its soft breaking. Starting from $\mathcal{N}=2$ $SU(2)$ gauge theory with $N_F=0$, the authors introduce soft SUSY breaking via a mass $M$ for the adjoint scalar in the $SU(2)$ multiplet. The soft breaking term is precisely mapped to the IR theory by embedding it in the $\mathcal{N}=2$ conserved stress-tensor multiplet \cite{Sohnius1979}. This mapping is inspired by \cite{Luty1999,Abel2011} but is a slightly more straightforward version, which is available in $\mathcal{N}=2$ theories. In the presence of the soft scalar mass $M\ll \Lambda$, the authors show that the origin becomes the global (and only) reliable minimum in the theory. A potential minimum associated with the monopole/dyon singularities turns out to be too far from them on the moduli space and so beyond the range of validity of the near-singularity EFT. The bulk of the paper explores the possible phase of the theory for $M\gg \Lambda$. Though it is impossible to reliably determine the IR phases in this case, every point on the $\mathcal{N}=2$ moduli space could become the global minimum while satisfying all 't Hooft anomalies in a non-trivial manner. The same is true even if the IR dynamics for $M\gg \Lambda$ continuously deforms the theory, for example by making the EFT near the monopole/dyon singularities more weakly coupled. In the latter case, the soft mass $M$ triggers monopole condensation which breaks $SU(1)_R\rightarrow U(1)_R$. The IR dynamics is then described by a $\mathbb{C}P^1$ model. 

    In most of \cite{Cordova2018}, the authors only consider the soft scalar mass $M$ and no gaugino masses. This is because their theory of interest is $N_F=2$ adjoint $SU(2)$ QCD, in which the adjoint fermions descend from the gauginos of the SUSY theory. In the absence of gaugino masses, the local minima associated with the monopole/dyon singularities turn out to be too far from them, which renders them unreliable. This is not the case when gaugino masses are turned on - for example in \cite{Seiberg1994,Seiberg1994a} and in our analysis the minima of the theory end up very close to the monopole/dyon singularities. This is the main difference between our analysis and \cite{Cordova2018}. We note that the latter paper does consider adding gaugino masses in sections 2.5 and 3.4 as a consistency check linking $N_F=2$ to $N_F=0,1$ adjoint $SU(2)$ QCD. If we take their analysis with hierarchical gaugino masses $0<m_1\ll m_2\ll\Lambda$ and set $M\rightarrow 0$, we get exactly our $N_F=0$ analysis.
    
    We note also the more recent \cite{DHoker2021,DHoker2022} which explored the multi-monopole points and the strong coupling region in general of $\mathcal{N}=2$ $SU(N)$ gauge theory. It would be interesting to add fundamental matter (as well as an $\mathcal{N}=1$ deformation and AMSB) to these theories to search for phase transitions at even more exotic values of $\theta$.

\end{itemize}

\section{Conclusion}
We performed a thorough analysis of the vacuum structure of $\mathcal{N}=2$ SQCD with gauge group $SU(2)$ and $N_F=0,1,2,3$ flavors, deformed to $\mathcal{N}=1$ by a mass $\mu$ and coupled to AMSB with SUSY breaking $m$. The hierarchy $m\ll \mu\ll \Lambda$ allows for a reliable analysis. {We find a branched structure of the potential generated by the condensation of the various monopoles/dyons that become massless on various points of the moduli space. Our results verify Witten's original picture where the $\theta$-dependence of the vacuum energy is generated through the mechanism responsible for confinement (rather than being a direct instanton contribution).
Our results indicate first order phase transitions between the different branches of the confining vacua as a function of $\theta$.} In particular, for $N_F=0$ there is a phase transition at $\theta=\pi$, while for $N_F=1$ the phase transitions are at $\theta=0,\pi$, for $N_F=2$ they are at $\theta=\frac{\pi}{2},\frac{3\pi}{2}$. As far as we know, this is the first time that first order phase transitions have been found at these values of $\theta$. Additionally, the $2\pi$ periodicity in $\theta$ is guaranteed non-trivially by the Witten effect.

The case $N_F=3$ is unique in several ways. First, our solution near the $k=2$ singularity is new in the literature. Furthermore, our results indicate first order phase transitions at $\theta=\frac{\pi}{4},\frac{3\pi}{4},\frac{5\pi}{4},\frac{7\pi}{4}$ between a chiral symmetry preserving vacuum near the origin, and a chiral symmetry breaking one near $u=\Lambda_3$. Finally, the dyons condensing near $u=\Lambda_3$ have magnetic charge 2, and so they hint at a topological $Z_2$ gauge theory in the IR.

The phase structure explored in this paper can be neatly explained by a mixed anomaly argument similar to the one presented in \cite{Gaiotto:2017yup}. In particular, we find that a combination of a discrete remnant of the UV r-symmetry $R$-symmetry and time reversal becomes unbroken -- exactly at the values of $\theta$ for which we find our 1st-order phase transitions. This 0-form symmetry has a mixed anomaly with a 1-form center symmetry of the UV theory, similar to \cite{Gaiotto:2017yup}. We leave the details of this analysis to upcoming work by some of the present authors.

The branched structure of the vacuum energy could have significant phenomenological consequences once $\theta$ becomes dynamical, namely when calculating axion potentials. In particular, The existence of degenerate vacua from different branches at multiple values of $\theta$ modifies the domain wall number of the IR theory. For example, if one only considered the $k=1$ vacuum for $N_F=2$, the domain wall number would seem to be $N_{DW}=1$. However, the existence of the $k=2$ vacuum leads to $N_{DW}=2$, with a domain wall between $\theta=0$ and $\theta=\pi$.

\appendix 

\section{Derivation of Local Sections and Prepotentials}\label{app:derivation}
For completeness, we present in this appendix the explicit expressions for the sections $(a_D,a)$ following the original calculation\footnote{Note that our solutions for $N_F=0,\ldots,3$ are S-dual to the ones in \cite{Ferrari1996}, except for our solution for the $k=2$ vacuum for $N_F=3$, which is our original derivation.} in \cite{Bilal1996,Ferrari1996}. Once the sections are known, the (magnetic) prepotential $\mathcal{F}$ is then found by substituting the ansatz \cite{Ito1996,Ito1996a}
\begin{eqnarray}
a^2(z)\left[A \log\left(\frac{a(z)}{\Lambda}\right)+\sum_{i=-1} B_i \left(\frac{a(z)}{\Lambda}\right)^i\right]\,,
\end{eqnarray}
and solving for $A,\,B_i$ by expanding the relation
\begin{eqnarray}
\mathcal{F}'\left[a(z)\right]=a_D(z)\,,
\end{eqnarray}
order-by-order in $\Delta z=z-z_{\rm singularity}$.

The sections $(a_D,a)$ are calculated as follows. For each number of flavors $0\leq N_F \leq 3$, we take the Seiberg-Witten curves \eqref{eq:SW0}-\eqref{eq:SWNF} and the monodromy matrices in tables~\ref{tab:NF0}-\ref{tab:NF3} as inputs - they are derived in \cite{Seiberg1994,Seiberg1994a} by holomorphy arguments.
From the Seiberg-Witten curves \eqref{eq:SW0}-\eqref{eq:SWNF} and the Seiberg-Witten differential \eqref{eq:SWdif} we extract the corresponding Picard-Fuchs equations \cite{Lerche1997,Ito1996,Ito1996a,Bilal1996,Ferrari1996},
\begin{eqnarray}\label{eq:PF}
4\Pi_{N_F}''(z)+f_{N_F}(z)\,\Pi_{N_F}(z)=0\,,~~~~z=\frac{u}{\Lambda^2_{N_F}}\,,
\end{eqnarray}
where $\Pi_{N_F}=a^{(N_F,k)}$ or $\Pi_{N_F}=a^{(N_F,k)}$, and
\begin{eqnarray}\label{eq:PFf}
\left(f_{0}(z),f_{1}(z),f_{2}(z),f_{3}(z)\right)=
\left(\frac{1}{z^2-1},\frac{z}{z^3+1},\frac{1}{z^2-1},\frac{1}{z(z-1)}\right)\,.
\end{eqnarray}
The solutions to \eqref{eq:PF} are expressed in terms of hypergeometric functions, and are fixed by the monodromies in~\ref{tab:NF0}-\ref{tab:NF3}. Note that these monodromies are in the weak coupling frame, and so we need to perform duality transformations to get to the local duality frame of each singularity, in which $a^{(N_F,k)}(u^{(N_F,k)})=0$. 
\begin{figure}[h]~
     \centering
     \begin{subfigure}[b]{0.45\textwidth}
         \centering
         \includegraphics[width=\textwidth]{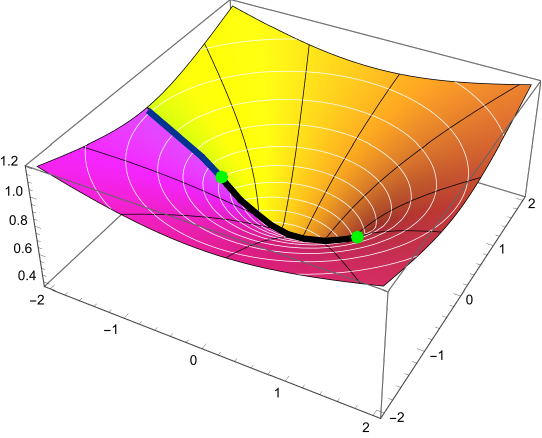}
         \caption{$a^{(0,1)}_D$}
         \label{fig:adNF0m1}
     \end{subfigure}
     \hfill
     \begin{subfigure}[b]{0.45\textwidth}
         \centering
         \includegraphics[width=\textwidth]{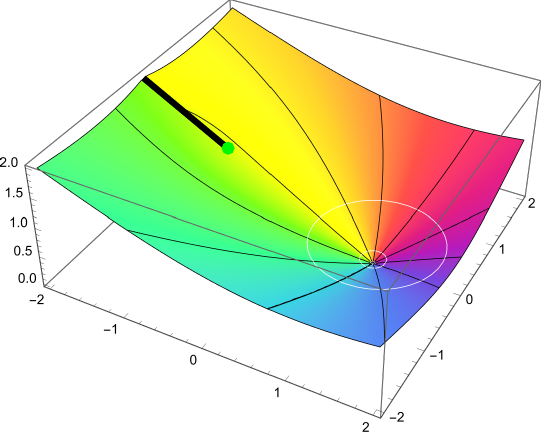}
         \caption{$a^{(0,1)}$}
         \label{fig:aNF0m1}
     \end{subfigure}
        \caption{$(a^{(0,1)}_D,a^{(0,1)})$ for $N_F=0$. The singularities in green are at $(u^0_1,u^0_2)/\Lambda^2_0=(-1,1)$. The black line represents a branch cut while the blue line represents the superposition of two branch cuts - one from $z=-1$ and one from $z=1$. The sections for $N_F=2$ are identical to these up to a factor of 2 in $a$.}
        \label{fig:aadNF0}
\end{figure}
\begin{figure}[h]~
     \centering
     \begin{subfigure}[b]{0.45\textwidth}
         \centering
         \includegraphics[width=\textwidth]{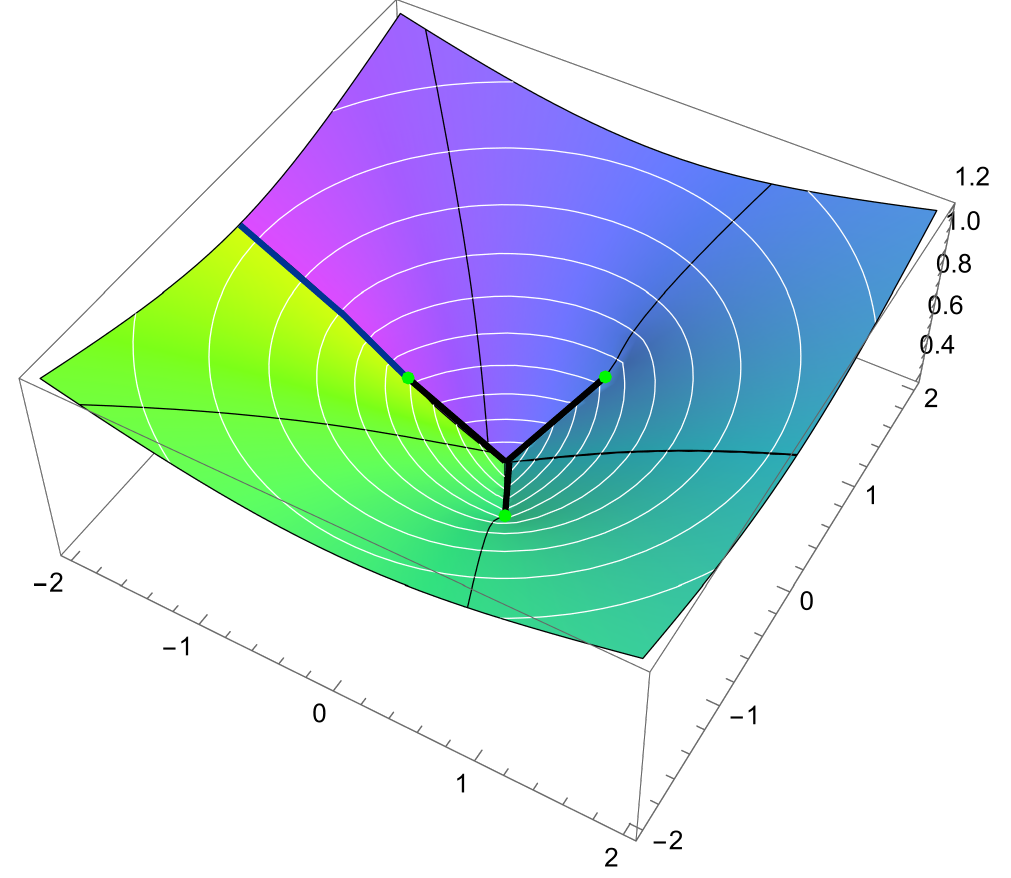}
         \caption{$a^{(1,3)}_D$}
         \label{fig:adNF1}
     \end{subfigure}
      \hfill
         \begin{subfigure}[b]{0.45\textwidth}
         \centering
         \includegraphics[width=\textwidth]{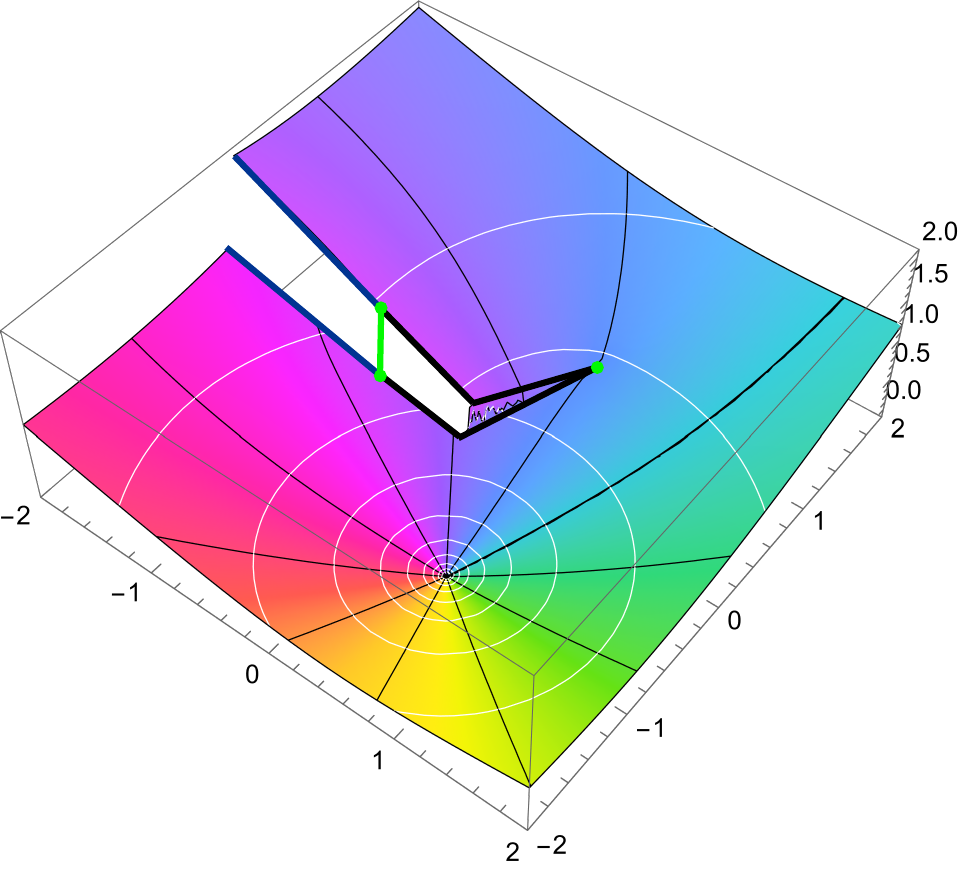}
         \caption{$a^{(1,3)}$}
         \label{fig:aNF1}
     \end{subfigure}
                \caption{$(a^{(1,3)}_D,a^{(1,3)})$ for $N_F=1$. The singularities in green are at $(u^{(1,1)},u^{(1,2)},u^{(1,3)})/\Lambda^2_1=(e^{-i\pi/3},-1,e^{i\pi/3})$. The black lines represent branch cuts while the blue line represents the superposition of two branch cuts - one from $z=-1$ and one from $z=0$. The green line is just an indication that $a$ is discontinuous across the branch cut, including at the $u/\Lambda_1=-1$ singularity.}
        \label{fig:aadNF1}
\end{figure}
\begin{figure}[h]~
     \centering
     \begin{subfigure}[b]{0.45\textwidth}
         \centering
         \includegraphics[width=\textwidth]{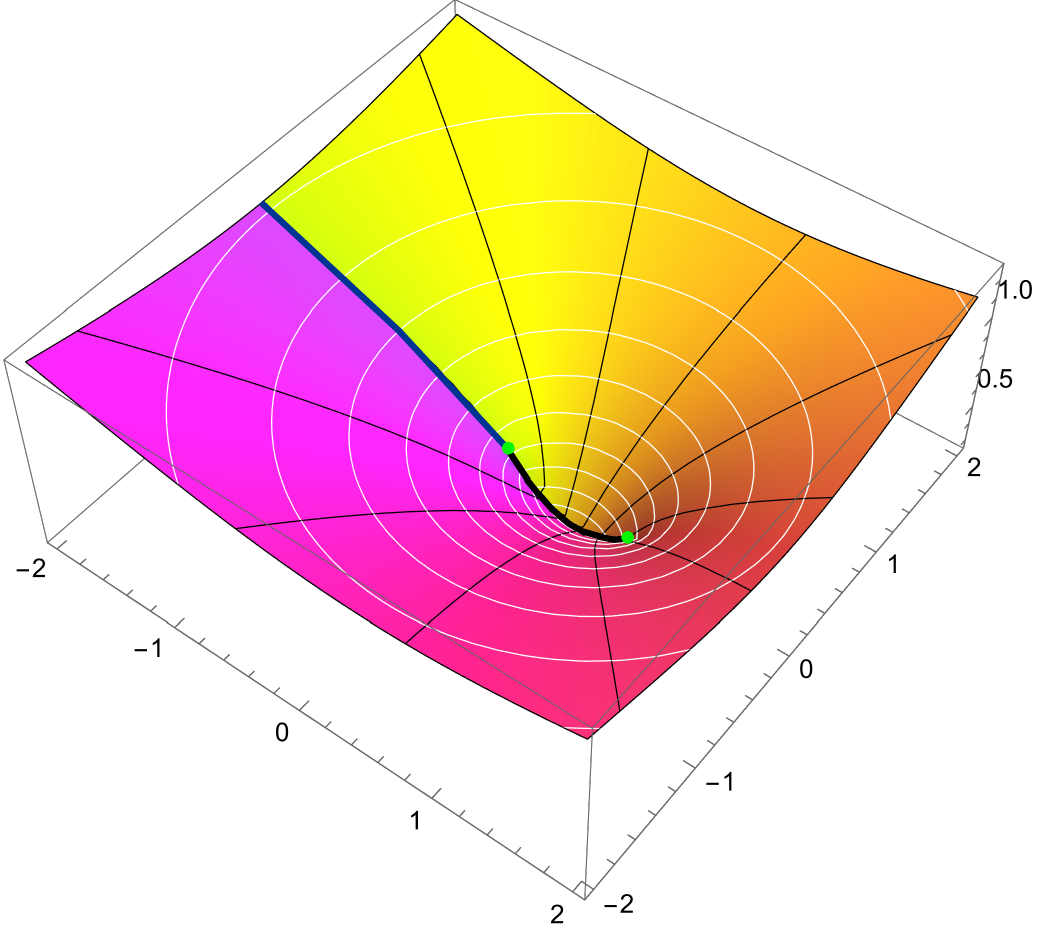}
         \caption{$a^{(3,1)}_D$}
     \end{subfigure}
     \vspace{10pt}
      \hfill
         \begin{subfigure}[b]{0.45\textwidth}
         \centering
         \includegraphics[width=\textwidth]{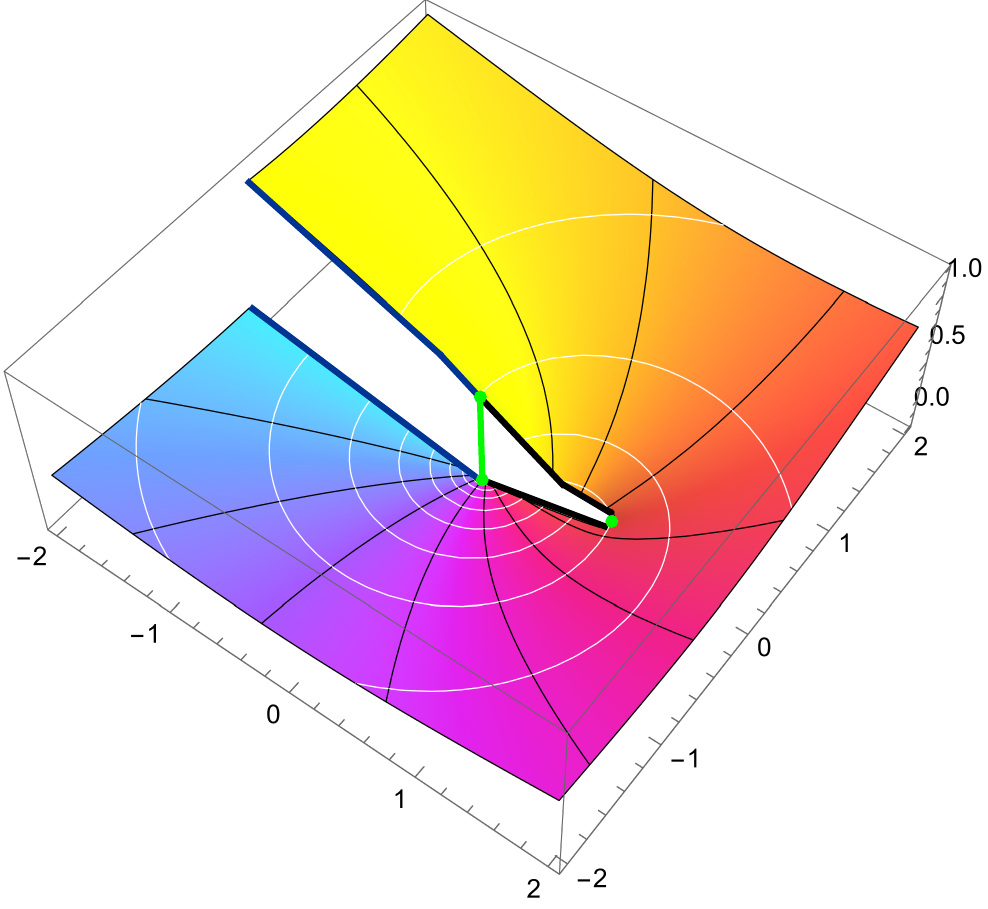}
         \caption{$a^{(3,1)}$}
     \end{subfigure}
          \begin{subfigure}[b]{0.45\textwidth}
         \centering
         \includegraphics[width=\textwidth]{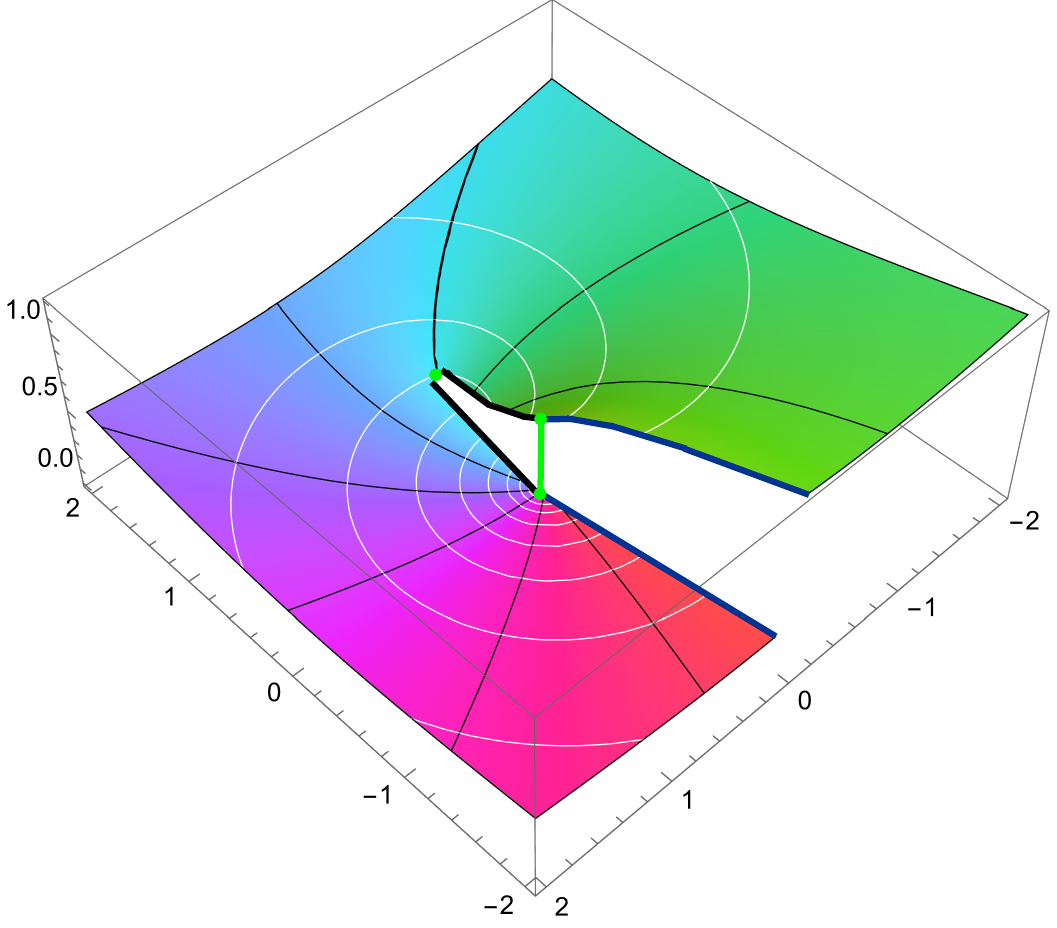}
         \caption{$a^{(3,2)}_D$}
     \end{subfigure}
      \hfill
         \begin{subfigure}[b]{0.45\textwidth}
         \centering
         \includegraphics[width=\textwidth]{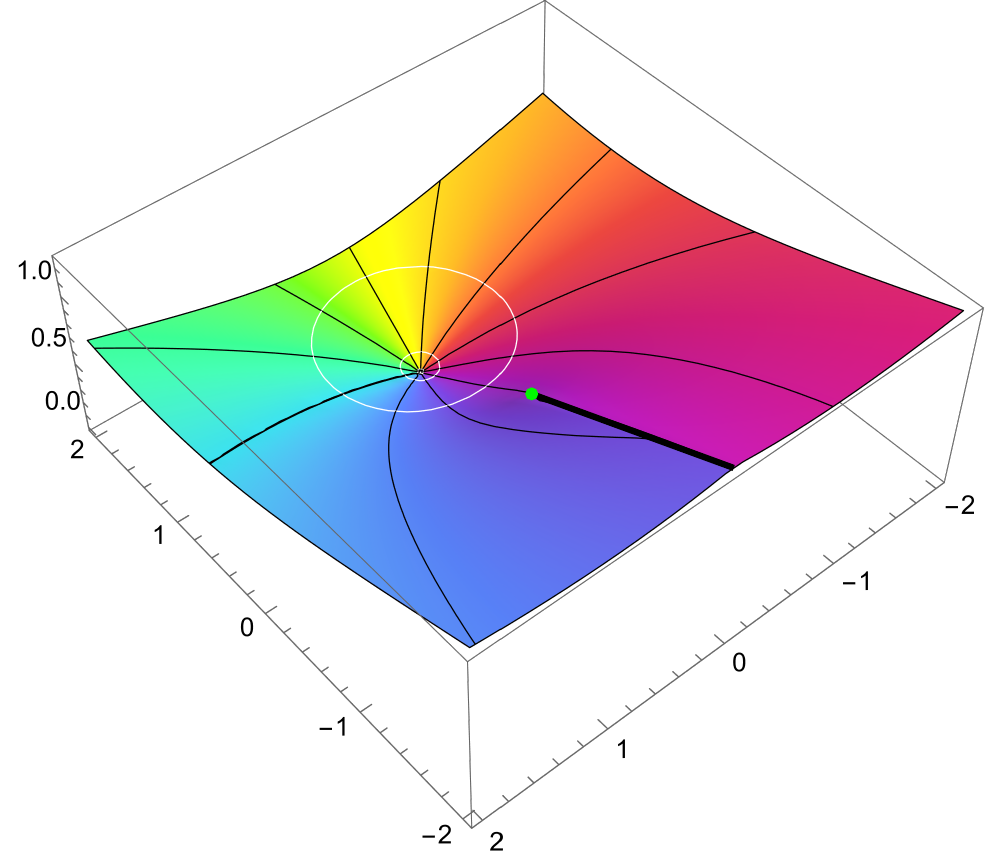}
         \caption{$a^{(3,2)}$}
     \end{subfigure}
           \caption{$(a^{(3,k)},a^{(3,k)})$ for $N_F=3$. The green singularities are at $(u^{(3,1)},u^{(3,2)})/\Lambda^2_3=\,(0,1)$. Note the different orientation of the $k=2$ plots in order to clearly see the jump on the branch cut. The black lines represent branch cuts while the blue line represents the superposition of two branch cuts - one from $z=0$ and one from $z=2$. The green line is just an indication of the discontinuity at the singularity $u=0$.}
        \label{fig:aadNF3}
\end{figure}
  
For $N_F=0$, the sections are \cite{Bilal1996,Ferrari1996}
\begin{eqnarray}\label{eq:solmon01}
\colmatt{a^{(0,1)}_D(z)\\a^{(0,1)}(z)}=\Lambda_0\,\colmatt{\sqrt{\frac{z+1}{2}}~{}_2F_1\left(-\tfrac{1}{2},\tfrac{1}{2},1;\tfrac{2}{z+1}\right)\\-i\frac{z-1}{2}~{}_2F_1\left(\tfrac{1}{2},\tfrac{1}{2},2;\tfrac{1-z}{2}\right)}\,,
\end{eqnarray}
where ${}_2F_1$ is the Gauss hypergeometric function. These solutions are depicted in Figure~\ref{fig:aadNF0}. It is straightforward yet tedious to check that these indeed have the right monodromies as in table~\ref{tab:NF0}, by analytically continuing $a^{(0,1)}$ across its branch cut to the left of $z=-1$, and $a^{(0,1)}_D$ across its branch cuts to the left and right of $z=-1$ and to the left of $z=1$. The section at the other singularity is then
\begin{eqnarray}\label{eq:solmon02}
\colmatt{a^{(0,2)}_D(z)\\a^{(0,2)}(z)}=-i\colmatt{a^{(0,1)}_D(-z)\\a^{(0,1)}(-z)}\,.
\end{eqnarray}
For $N_F=1$ we first define two auxiliary functions \cite{Bilal1996},
\begin{eqnarray}\label{eq:solmon1}
\colmatt{b^{(1)}(z)\\c^{(1)}(z)}=\Lambda_1\,\colmatt{\frac{z^3+1}{3\sqrt{8}}~{}_2F_1\left(\tfrac{5}{6},\tfrac{5}{6},2;z^3+1\right)\\\sqrt{\frac{z}{2}}~{}_2F_1\left(-\tfrac{1}{6},\tfrac{1}{6},1;-\tfrac{1}{z^3}\right)}\,.
\end{eqnarray}
We can now define
\begin{eqnarray}
a^{(1,3)}_D(z)&=&c^{(1)}(z),\nonumber\\[5pt]
a^{(1,3)}(z)&=&-\begin{cases}
-b^{(1)}(z)+c^{(1)}(z)&-\pi<\rm arg(z)\leq-\tfrac{2}{3}\pi\\
\omega_3\,b^{(1)}(z)&-\tfrac{2}{3}\pi<\rm arg(z)\leq0\\
-\omega^2_3\,b^{(1)}(z)-c^{(1)}(z)&~~~~0<\rm arg(z)\leq\tfrac{2}{3}\pi\\
b^{(1)}(z)-2c^{(1)}(z)&~~\tfrac{2}{3}\pi<\rm arg(z)\leq\pi
\end{cases}\,.
\end{eqnarray}
These solutions are depicted in Figure~\ref{fig:aadNF1}. One can check that these indeed have the right monodromies as in table~\ref{tab:NF2}, by analytically continuing $a^{(1,3)}$ across its branch cuts from $z=0$ to $z=e^{i\pi/3}$, from $z=-1$ to $z=0$ and to the left of $z=-1$. Similarly, one has to analytically continue $a^{(1,3)}_D$ across the same branch cuts as $a^{(1,3)}$, as well as from $z=0$ to $z=e^{-i\pi/3}$.

For $N_F=2$, the solutions are
\begin{eqnarray}\label{eq:solmon2}
\colmatt{a^{(2,1)}_D(z)\\a^{(2,1)}(z)}=\frac{\Lambda_2}{\Lambda_0}\colmatt{\tfrac{1}{2}a^{(0,1)}_D(z)\\a^{(0,1)}(z)}\,.
\end{eqnarray}
Finally, for $N_F=3$ we define the auxiliary functions
\begin{eqnarray}\label{eq:solmon3}
\colmatt{b^{(3)}(z)\\c^{(3)}(z)}=\Lambda_3\,\colmatt{i\frac{z-1}{2^{5/2}}~{}_2F_1\left(\tfrac{1}{2},\tfrac{1}{2},2;1-z\right)\\\sqrt{\frac{z}{2}}~{}_2F_1\left(-\tfrac{1}{2},\tfrac{1}{2},1;\tfrac{1}{z}\right)}\,.
\end{eqnarray}
In terms of these, the $N_F=3$ sections are given by
\begin{eqnarray}
\colmatt{a^{(3,1)}_D(z)\\a^{(3,1)}(z)}&=&\colmatt{c^{(3)}(z)\\-b^{(3)}(z)+\tfrac{1}{2}c^{(3)}(z)},\nonumber\\[5pt]
\colmatt{a^{(3,2)}_D(z)\\a^{(3,2)}(z)}&=&\colmatt{-b^{(3)}(z)-\tfrac{1}{2}c^{(3)}(z)\\2b^{(3)}(z)}
\,.
\end{eqnarray}
These solutions are depicted in Figure~\ref{fig:aadNF3}.They have the correct monodromies around the singularities at $z=0$ and $z=1$, as can be checked by analytically continuing through the branch cuts between $z=0$ and $z=1$ and left of $z=0$. 

\section*{Acknowledgments}

 OT would like to thank Ofer Aharony, Michael Dine, John Terning and Shimon Yankielowicz for useful discussions. 
TK is grateful to Yuji Tachikawa and Yoshio Kikukawa for helpful discussions. He also would like to thank Shigeki Sugimoto and Kazuya Yonekura for correspondence. CC thanks the Kavli IPMU in Tokyo for its hospitality while this paper was prepared. 
CC is supported in part by the NSF grant PHY-2309456 and in part by the US-Israeli BSF grant 2016153. 
OT is supported by the ISF grant No.~3533/24 and by the NSF-BSF grant No.~2022713. HM is the Hamamatsu Professor at the Kavli IPMU in Tokyo. HM was also supported in part by the NSF grant
PHY-2210390, by the DOE under grant DE-AC02-05CH11231, by the US-Israeli BSF grant 2022287, by the JSPS Grant-in-Aid for
Scientific Research JP23K03382, MEXT Grant-in-Aid for Transformative Research Areas (A)
JP20H05850, JP20A203, by WPI, MEXT, Japan, and Hamamatsu Photonics, K.K.  CC and HM  also thank the Munich Institute for Astro-, Particle and BioPhysics (MIAPbP - funded by the DFG under Germany's Excellence Strategy EXC-2094-390783311) for its hospitality while this paper was being concluded.

\bibliographystyle{utcaps_mod}
\providecommand{\href}[2]{#2}\begingroup\raggedright\endgroup

\end{document}